\documentclass{jpp}
\usepackage{graphicx}
\usepackage{epstopdf, epsfig}
\usepackage{hyperref}
\usepackage{amssymb}
\usepackage{amsmath}
\usepackage{xcolor}

\renewcommand\vec\mathbf

\shorttitle{Wall ion velocity distribution in a shallow-angle magnetic field}
\shortauthor{A. Geraldini}

\title{Large gyro-orbit model of ion velocity distribution in plasma near a wall in a grazing-angle magnetic field}

\author{Alessandro Geraldini\aff{1}
  \corresp{\email{ale.gerald@gmail.com}}}
\affiliation{\aff{1}Institute for Research in Electronics and Applied Physics, University of Maryland, College Park, MD 20742, USA}
\begin{document}

\maketitle

\begin{abstract}
A model is presented for the ion distribution function in a plasma at a solid target with a magnetic field $\vec{B}$ inclined at a small angle, $ \alpha \ll 1$ (in radians), to the target. 
Adiabatic electrons are assumed, requiring $\alpha\gg\sqrt{Zm_{\rm e}/m_{\rm i}} $ where $m_{\rm e}$ and $m_{\rm i}$ are the electron and ion mass respectively, and $Z$ is the charge state of the ion.
An electric field $\vec{E}$ is present to repel electrons, and so the characteristic size of the electrostatic potential $\phi$ is set by the electron temperature $T_{\rm e}$, $e\phi \sim T_{\rm e}$, where $e$ is the proton charge. 
An asymptotic scale separation between the Debye length, $\lambda_{\rm D} = \sqrt{\epsilon_0 T_{\text{e}} / e^2 n_{\text{e}} } $, the ion sound gyroradius $\rho_{\rm s} = \sqrt{ m_{\rm i} ( ZT_{\rm e} + T_{\rm i} ) } / (ZeB)$, and the size of the collisional region $d_{\rm c} = \alpha \lambda_{\rm mfp}$ is assumed, $\lambda_{\rm D} \ll \rho_{\rm s} \ll d_{\rm c}$.
Here $\epsilon_0$ is the permittivity of free space, $n_{\rm e}$ is the electron density, $T_{\rm i}$ is the ion temperature, $B= |\vec{B}|$ and $\lambda_{\rm mfp}$ is the collisional mean free path of an ion.
The form of the ion distribution function is assumed at distances $x$ from the wall such that $\rho_{\rm s} \ll x \ll d_{\rm c}$, i.e. collisions are not treated.
A self-consistent solution of the electrostatic potential for $x \sim \rho_{\rm s}$ is required to solve for the quasi-periodic ion trajectories and for the ion distribution function at the target.
The large gyro-orbit model presented here allows to bypass the numerical solution of $\phi (x)$ and results in an analytical expression for the ion distribution function at the target.  
It assumes that $\tau=T_{\rm i}/(ZT_{\rm e})\gg 1$, and ignores the electric force on the quasi-periodic ion trajectory until close to the target.
For $\tau \gtrsim 1$, the model provides an extremely fast approximation to energy-angle distributions of ions at the target.
These can be used to make sputtering predictions.
\end{abstract}

\section{Introduction}

When plasma is in contact with a solid surface --- such as in fusion experiments \citep{Stangeby-book}, Hall thrusters \citep{Boeuf-2017}, plasma probes \citep{Hutchinson-book}, magnetic filters \citep{Anders-1995-filters}, and orbiting spacecraft \citep{Hastings-1995} --- the resulting interaction affects both the plasma and the surface.
Among the many plasma-surface interaction processes, one that is of particular concern is sputtering, where an ion from the plasma reaches the surface material and knocks an atom off the surface. 
Ionization of sputtered atoms in the plasma produces impurities, thus altering the plasma.
Moreover, in the long run sputtering causes erosion of the surface material. 
The amount of sputtering depends on a wide variety of factors, including surface material, surface roughness, plasma conditions and velocity distributions of particles striking the target \citep{Krashenninikov-2017, Cohen-Ryutov-1998-roughness, Drobny-2017, Khaziev-Curreli-2015, Siddiqui-Hershkowitz-2016, Lasa-2020}. 

\begin{figure} 
\centering
\includegraphics[width=0.8\textwidth]{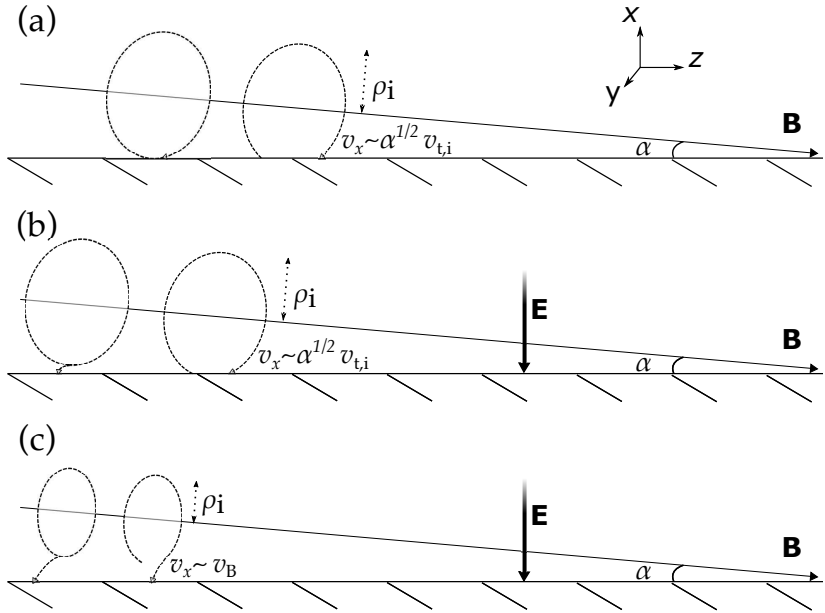}
\caption{Cartoons of ion gyro-orbits, whose gyro-radius is $\rho_{\rm i}$, reaching the target when the angle between the magnetic field $\vec{B}$ and the target is small, $\alpha \ll 1$. The axes $(x,y,z)$ are labelled. (a) With no normal electric field, the circular orbit moves closer to the target by $\alpha \rho_{\rm i}$ after a gyroperiod and thus the normal velocity of an ion at the target is $v_x \sim \sqrt{\alpha} v_{\rm t,i}$. % and the ion strikes the target at an angle $\theta \sim \sqrt{\alpha}$ (in radians). 
(b,c) With the magnetic presheath and Debye sheath electric field $\vec{E}$, ions are accelerated to $v_x \sim \sqrt{\alpha  v_{\rm t,i}^2 + v_{\rm B}^2 }$. }
\label{fig-ion}
\end{figure}

In this paper, we focus on the calculation of the distribution function of plasma ions striking the solid surface.
We consider the target surface --- or wall --- to be smooth, planar and absorbing all incident particles.
We consider a plasma magnetized by a uniform magnetic field $\vec{B}$, with one ion species.
The angle between the magnetic field and the wall is taken to be small, $\alpha \ll 1$ (measured in radians unless otherwise indicated).
This situation is particularly relevant in fusion plasmas, where divertors are designed so that the angle between incident magnetic field lines and the target surface is as small as possible. 
We define a set of right-handed cartesian axes $(x,y,z)$ where $x$ measures the distance from the wall, $z$ measure displacements in the direction tangential to the wall, such that the magnetic field is in the $x$-$z$ plane, and $y$ measures displacements in the remaining direction. 
The axes are shown on the top-right of figure \ref{fig-ion}.
For simplicity, we assume no gradients tangential to the wall.
Thus, the only gradients are in the $x$ direction.

The standard picture of the plasma-wall boundary is as follows.
Close to the wall, there is a thin positively charged layer called \emph{Debye sheath}, with a characteristic size of a few Debye lengths $\lambda_{\text{D}} = \sqrt{\epsilon_0 T_{\text{e}} / e^2 n_{\text{e}} } $, where a strong electric field $\vec{E} = - \nabla \phi$ directed towards the target is present to repel electrons \citep{Riemann-review, Hershkowitz-2005, Baalrud-2020}.
Here, $e$ is the proton charge, $n_{\text{e}}$ is the number density of the electrons, $\epsilon_0$ is the permittivity of free space, $T_{\text{e}}$ is the temperature of the electrons, and $\phi (x)$ is the electrostatic potential as a function of the distance from the wall.
The purpose of the electric field is to achieve a steady state with comparable (or, in ambipolar conditions, equal) fluxes of ions and electrons to the wall.
The size of the electrostatic potential drop necessary to repel electrons is $| \phi |\sim T_{\rm e} / e $.
The kinetic energy gained by an ion of charge $Ze$ in such a potential is $Z e |\phi | \sim ZT_{\rm e}$.
Hence, the parameter
\begin{align}
\tau = \frac{T_{\rm i}}{ZT_{\rm e}} \rm ,
\end{align}
where $T_{\rm i}$ is the ion temperature, is a measure of the ratio of ion thermal energy divided by ion kinetic energy gained from the electric field.
At the edge of a fusion device one often finds $\tau \gtrsim 1$ \citep{Mosetto-2015}.
Poisson's equation, 
\begin{align} \label{Poisson}
\varepsilon_0 \phi'' (x) = Zen_{\rm i}(x) - en_{\rm e}(x) \rm ,
\end{align}
relates the charge separation to the electrostatic potential in the Debye sheath, where $x \sim \lambda_{\rm D}$.
Here a prime denotes differentiation with respect to the argument, in this case $x$, of the function.
At distances from the wall comparable to the ion sound gyroradius, $\rho_{\text{s}} $, the ion population is depleted due to a combination of ion gyro-orbit losses and acceleration of ions by the electric field, as schematically shown in figure \ref{fig-ion}.
Here, $\rho_{\rm s} = c_{\rm s} / \Omega $, where $c_{\rm s} = \sqrt{ ( ZT_{\rm e} + T_{\rm i} ) / m_{\rm i}}$ is the ion sound speed, $\Omega = ZeB / m_{\rm i}$ is the ion gyrofrequency, $B = |\vec{B}|$ and $m_{\rm i}$ is the ion mass.
Since typically $\lambda_{\rm D} \ll \rho_{\rm s}$, the region $x \sim \rho_{\rm s}$ can be assumed to be quasineutral,
\begin{align} \label{quasineutrality}
Zn_{\rm i}(x) \simeq n_{\rm e} (x) \rm ,
\end{align}
and is referred to as \emph{magnetic presheath} (and sometimes as Chodura sheath). 
A substantial fraction of the electrostatic potential drop between the plasma and the wall must occur in the magnetic presheath, as an electric field is necessary to adjust the electron and ion densities such that (\ref{quasineutrality}) is preserved.
At typically even larger distances from the target, $d_{\rm c} \gg \rho_{\rm s}$, ions tend to collide with neutrals or other ions before reaching the target.
Thus, the magnetic presheath and Debye sheath can be assumed to be collisionless.
In this paper, the form of the ion distribution function in the region $\rho_{\rm s} \ll x \ll d_{\rm c}$ is assumed.
This region is known as the magnetic presheath entrance.

Several distinct approaches may be used to calculate the velocity distributions of ions reaching the target.
An approach that describes all the phenomena at play close to the wall, including the effect of the collisional layer, is to numerically solve the kinetic Vlasov equation for the ions and electrons self-consistently with the Poisson equation for the electrostatic potential \citep{Coulette-Manfredi-2016}.
An alternative, equally complete, approach is the particle-in-cell (PIC) method \citep{Tskhakaya-2003, Khaziev-Curreli-2015}.
Both the Vlasov and the PIC approaches offer the most complete description of the plasma, but can be computationally expensive.
Simplifying models can offer more immediate calculations.
For example, taking into account gyro-orbit losses at the wall, but ignoring the electric field, one can solve for distribution functions at the wall analytically, assuming an incoming Maxwellian \citep{Parks-Lippmann-1994} or more refined boundary conditions \citep{Gunn-2017}.
However, in neglecting the electric field this model assumes that some ions can reach the target travelling tangentially\footnote{One could add the kinetic energy gain of an ion in the Debye sheath ad hoc. However, the resulting velocity distributions would vastly overestimate the energy going into the normal component of the ion velocity and the angle of impact of ions with the target.}, as the left ion in figure \ref{fig-ion}(a) does.
By introducing an ad hoc analytical electrostatic potential function close to the wall to model the effect of gyro-orbit distortion, \cite{Borodkina-2016} numerically solved for ion trajectories near the target.
The authors found a substantial effect on erosion coefficients, as was also suggested by \cite{Siddiqui-Hershkowitz-2016}.
\cite{Daube-Riemann-1999} obtained self-consistent solutions of the electrostatic potential and ion distribution function in a magnetic presheath by considering charge exchange collisions with cold neutrals.
They calculated the ion density as an integral over characteristics originating at the last collision event.
The resulting ion distribution functions exhibit an interesting and involved structure with singularities, which are 
expected to be smeared out by unstable ion cyclotron modes \citep{Daube-1998} and finite neutral temperature. 
\cite{Tskhakaya-2014} analysed the plasma-wall boundary layers using an asymptotic scale separation and an asymptotic expansion in $\alpha \ll 1$.
They considered the ion gyro-orbits to have zero spatial extent, but retained all other kinetic effects.
In \cite{Geraldini-2017}, the full approximately periodic ion trajectories in the collisionless magnetic presheath were solved using an expansion in $\alpha \ll 1$.
This expansion leads to the presence of an adiabatic invariant, as first described by \cite{Cohen-Ryutov-1998}.
A numerical scheme to efficiently calculate the self-consistent electrostatic potential was developed by \cite{Geraldini-2018}.
The final open piece of the ion trajectory near the wall was included in the ion density calculation. 
Velocity distributions of ions reaching the Debye sheath, consistent with a quasineutral magnetic presheath, were thus obtained.
While this treatment applies only to grazing angles, it provides an efficient way to solve self-consistently for the effect of the electric field on ion trajectories in the collisionless magnetic presheath.

In this paper a large gyro-orbit model for the ion distribution function at the target is developed.
The full solution of the self-consistent electrostatic potential is bypassed.
Instead, the electrostatic potential is assumed to distort ion gyro-orbits only just before ions reach the Debye sheath.
This assumption is expected to be more accurate for large gyro-orbits, $\tau \gg 1$.
The model results are compared with distribution functions obtained using the full self-consistent electrostatic potential solution in the magnetic presheath, with good qualitative agreeement for $\tau \gtrsim 1$. 
The agreement between the two methods is better at larger values of $\tau$, as expected.

The rest of the paper is structured as follows.
In section~\ref{sec-orderings}, the orderings assumed in this work are presented and discussed.
In section~\ref{sec-electrons}, the electron model is introduced.
In section \ref{sec-trajectories} ion trajectories in the collisionless magnetic presheath and Debye sheath regions are analyzed.
Expressions for the velocity distributions of ions reaching the Debye sheath and of ions striking the target are obtained in section~\ref{sec-IVDF}.
These expressions depend on the full electrostatic potential solution in the magnetic presheath, $\phi (x)$.
The trajectories of ions in large gyro-orbits, for $\tau \gg 1$, are analyzed in section \ref{sec-VDMP-model}.
From this analysis, a model for the ion velocity distribution at the target is developed.
In section~\ref{sec-numerical} ion distribution functions obtained from the large gyro-orbit model are compared to ones obtained from the full self-consistent electrostatic potential solution $\phi (x)$ in the magnetic presheath.
Finally, in section~\ref{sec-conclusions}, the results of the paper are summarized.

\section{Orderings} \label{sec-orderings}

As mentioned in the introduction, the typical electrostatic potential variation across the magnetic presheath and Debye sheath is ordered as $ |\phi | \sim T_{\rm e} / e $.
Hence, the kinetic energy tranferred by the electric field to an ion of charge $Ze$ is $Z e \phi \sim ZT_{\rm e}$ and the characteristic speed of an ion due to the energy gained from the electric field is the Bohm velocity $v_{\rm B} = \sqrt{ZT_{\rm e} / m_{\rm i}}$.
The thermal energy of an ion is $T_{\rm i}$ and the thermal speed of an ion is $v_{\rm t,i} = \sqrt{2T_{\rm i} / m_{\rm i}}$.
Adding together the contributions to the energy, the typical kinetic energy of an ion is $ZT_{\rm e} + T_{\rm i}$.  % = \sqrt{2T_{\rm i} / m_{\rm i}}$
The ion velocity, denoted $\vec{v} = (v_x, v_y, v_z)$ where $v_k$ is the velocity component in the $k$th direction, is therefore ordered such that $|\vec{v}|  \sim  \sqrt{(ZT_{\rm e} + T_{\rm i} ) / m_{\rm i}} = c_{\rm s}$.

The presence of ion gyro-orbits and the grazing angle of the magnetic field with the target modify the ordering for $v_x$ at the target as follows. 
Consider a circular ion gyro-orbit with no electric field, as shown in figure \ref{fig-ion}(a).
The component of the velocity parallel to the magnetic field is denoted $v_{\parallel}$ and the magnitude of the gyrating component of the velocity is denoted $v_{\perp}$.
The gyrophase angle of the ion is denoted $\varphi$.
In the small-angle approximation, $\sin \alpha \simeq \alpha$, $\cos \alpha \simeq 1$, and the component of the velocity normal to the wall is given by $v_x \simeq v_{\perp}  \sin \varphi - \alpha v_{\parallel} $.
If the gyro-orbit almost touches the wall ($x \rightarrow 0$) tangentially at a time $t=0$, the distance from the wall at a later time $t$ is $x \simeq (v_{\perp} / \Omega ) \left( 1  - \cos \varphi \right) - \alpha v_{\parallel} t $.
After a full gyro-period $2\pi / \Omega$, the orbit has drifted a little closer to the wall.
Therefore, the gyrophase angle corresponding to $x=0$ is no longer $\varphi = 0$, yet it has only changed by a small amount.
Solving for $x=0$ at $t = 2\pi / \Omega$ with $1 - \cos \varphi \simeq \varphi^2 / 2$ gives $\varphi \simeq \sqrt{ 4 \pi \alpha v_{\parallel} / v_{\perp} }$, and thus $v_x \simeq  - \sqrt{4 \pi \alpha v_{\parallel} v_{\perp} } $ \citep{Cohen-Ryutov-1998}.
The piece of $v_x$ equal to $-\alpha v_{\parallel}$ is smaller by a factor of $\sqrt{  \alpha v_{\parallel} / ( 4\pi v_{\perp} ) }$, and can be neglected.
Thus, the gyrophase dependence of ions reaching the target gives rise to an interval in allowed values of normal kinetic energy, $0 \leqslant v_x^2 / 2 < 2\pi \alpha v_{\parallel} v_{\perp} $.
The electric field, however, can still accelerate the ions by transferring an energy $\sim ZT_{\rm e}$ to the normal component of the velocity, as depicted schematically in figure \ref{fig-ion}(b-c).
Note that this additional acceleration towards the target is not obvious. 
It only happens because, as we will see, the electric field close to the target is sufficiently inhomogenous ($|\phi''(x)|$ is sufficiently large) that it overcomes the magnetic force pulling the ion back away from the target.
Combining these two contributions to the normal kinetic energy gives $v_x^2 / 2 \sim ZT_{\rm e} + \alpha T_{\rm i}$. 
The velocity of the ion at the target therefore satisfies $v_x \sim v_{\rm B} \sqrt{ 1+\alpha \tau }$ and $v_y \sim v_z \sim c_{\rm s}$.

As was discussed in the introduction, the Debye sheath, the magnetic presheath and the collisional region are assumed to satisfy the scale separation $\lambda_{\rm D} \ll \rho_{\rm s} \ll d_{\rm c}$.
At distances $x \sim d_{\rm c} \gg \rho_{\rm s}$, the ion motion is restricted along a field line.
Therefore, the size of the collisional region can be expressed as $d_{\rm c} \sim \alpha \lambda_{\rm mfp}$, where $\lambda_{\rm mfp}$ is the mean free path of an ion near the target.
It follows that the angle $\alpha$ must satisfy $\alpha \gg \rho_{\rm s} / \lambda_{\rm mfp}$ in order for $\rho_{\rm s } \ll d_{\rm c}$ to be valid.

In order to simplify the treatment of the electrons, the electron gyroradius $\rho_{\rm e} =  \sqrt{2 m_{\rm e} T_{\rm e}} / (eB)$ is assumed to be much smaller than the Debye length, such that $\rho_{\rm e} \ll \lambda_{\rm D}$ \citep{Stangeby-2012, Loizu-2012}.
Being tightly bound to the magnetic field lines, electrons have to travel along the magnetic field in order to reach the wall.
The typical speed of an electron is the electron thermal speed, $ v_{\rm t,e} = \sqrt{2T_{\rm e}/m_{\rm e}}$.
Conversely, the typical ion velocity close to the wall is $\sim v_{\rm B} \sqrt{ 1+\alpha \tau }$ towards the wall.  
When unopposed by an electric field, the electrons reach the wall much more quickly than the ions provided that $\alpha v_{\rm t,e} \gg v_{\rm B} \sqrt{ 1+\alpha \tau }$, or $\sqrt{ (1+\alpha \tau ) Zm_{\rm e} / m_{\rm i} }  \ll \alpha$.
For $\alpha \tau \lesssim 1$, the ordering $\alpha \gg \sqrt{ Z m_{\rm e} / m_{\rm i} }$ emerges.
For $\alpha \tau \gg 1$, the ordering $\alpha \gg m_{\rm e} \tau Z / m_{\rm i}  $ emerges instead.
Putting these last two orderings together gives $ 1/ \alpha \ll \tau \ll \alpha m_{\rm i} / m_{\rm e} Z $, which can only be satisfied if, again, $\alpha \gg \sqrt{ Z m_{\rm e} / m_{\rm i} }$.
To summarize, for $\alpha \gg \sqrt{ Z m_{\rm e} / m_{\rm i} }$ the electrons reach the target much more quickly than the ions.
An electric field must therefore set up to repel most of the electrons from the target.

Summarizing the orderings of this work, the physical length scales satisfy
\begin{align} \label{ordering-scales}
\rho_{\rm e} \ll \lambda_{\rm D} \ll \rho_{\rm s} \ll d_{\rm c}  \rm .
\end{align}
The angle and mass ratio satisfy
\begin{align} \label{ordering-angle}
\sqrt{\frac{ Zm_{\rm e}}{m_{\rm i}} }  \ll \alpha \ll 1 \rm .
\end{align}
The validity of these orderings is examined for a current fusion experiment such as JET.
In a Deuterium plasma, the angle obtained from the square root of mass ratio is $\sqrt{Zm_e/m_i} \approx 0.02~ \rm rad \sim 1^{\circ}$.
From \cite{Militello-Fundamenski-2011}, we estimate for JET: $B \sim 2~ \rm T$, $T_{\rm e} \sim T_{\rm i} \sim 30~ \rm eV$, $n_{\rm e} \sim n_{\rm i} \sim 10^{19} ~\rm m^{-3}$, giving $\rho_{\rm s} \sim 1~ \rm mm$, $\lambda_{\rm D} \sim \rho_{\rm e} \sim 0.01 ~\rm mm $ and $\alpha  \approx 0.07 \text{ rad} \approx 4^{\circ}$.
Since, of all the orderings in this paper, $\sqrt{Zm_{\rm e} / m_{\rm i} } \ll \alpha $ and $ \rho_{\rm e} \ll \lambda_{\rm D}$ are the least well-satisfied in fusion devices, it will be necessary to study in more detail the effect of electron inertia and gyroradius.

\section{Electron model} \label{sec-electrons}

In this work, Maxwellian electrons are assumed to enter the magnetic presheath.
We proceed to obtain the relationship between the electron current to the wall and the electrostatic potential at the wall.
We also derive, using the ordering (\ref{ordering-angle}), the Boltzmann expression for the electron density in the magnetic presheath. 

\begin{figure} 
\centering
\includegraphics[width=0.6\textwidth]{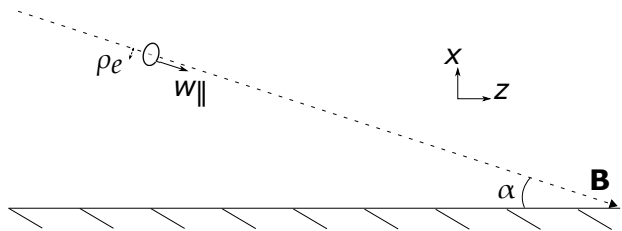}
\caption{Cartoon of an electron gyro-orbit, whose gyroradius is $\rho_{\rm e}$, streaming towards the wall along the magnetic field $\vec{B}$ with velocity $w_{\parallel}$.}
\label{fig-electron}
\end{figure}

According to (\ref{ordering-scales}), the electron gyroradius is so small that electrons are essentially tied to the magnetic field line, as shown in figure \ref{fig-electron}.
The electrons stream parallel to the magnetic field with a velocity given by $w_{\parallel}$.
At the very small length scale $\rho_{\rm e} \ll \lambda_{\rm D}$, the electron gyro-motion is unaffected.
The electron distribution function \emph{entering} (that is, for $w_{\parallel} > 0$) the magnetic presheath is assumed to be a half-Maxwellian,
\begin{align}
g_{\rm MPE} ( w_{\parallel} ) = Z\bar{n}_{\rm MPE}  \left( \frac{ m_{e}}{2\pi  T_e} \right)^{1/2} \exp \left(- \frac{m_e w_{\parallel}^2}{2T_e} \right) \text{ for } w_{\parallel} > 0 \text{,}
\end{align}
with density denoted as $Zn_{\rm MPE}$,
\begin{align}
Zn_{\rm MPE}= \int_{-\infty}^{\infty} g_{\rm MPE} ( w_{\parallel} ) dw_{\parallel} \text{.}
\end{align}
We set the zero of the electrostatic potential to be at the magnetic presheath, $\phi_{\rm MPE} = 0$.
Assuming the electrostatic potential to be a monotonically increasing function of $x$, the number of electrons that enter the magnetic presheath and come back out of it depends on the electrostatic potential at the wall relative to the magnetic presheath entrance, denoted $\phi_{\rm W} = \phi (0) < 0$.
Therefore, the constant $\bar{n}_{\rm MPE}$ depends on $n_{\rm MPE}$ and $\phi_{\rm W} $.

In the magnetic presheath and Debye sheath, the component of the electron velocity parallel to the magnetic field as a function of $x$ is obtained by energy conservation,
\begin{align} \label{vpar-electron}
w_{\parallel} = \sigma \sqrt{w_{\parallel \rm MPE}^2 + \frac{2e\phi(x)}{m_{e}}} \text{.}
\end{align}
Here, $w_{\parallel \rm MPE}$ is the electron velocity at the magnetic presheath entrance.
The $\vec{E}\times \vec{B}$ and gyration velocities of an electron remain unaffected by electrostatic potential variations as these have a much longer scale length than the electron gyroradius, $\lambda_{\rm D} \gg \rho_{\rm e} $.
In (\ref{vpar-electron}), $\sigma = \pm 1$ for those electrons reflected before reaching the wall and $\sigma = 1$ for those electrons that are not reflected.
At $x = 0$ the electron velocity is zero if $w_{\parallel \rm MPE}^2   = - 2e\phi_{\rm W} / m_{e}$.
Hence, reflected electrons satisfy
\begin{align}
w_{\parallel \rm MPE}^2   < - \frac{2e \phi_{\rm W}}{m_{e}} \text{,}
\end{align}
as they cannot reach $x=0$.
Therefore, the full electron distribution function at the magnetic presheath entrance is
\begin{align} \label{g-electron}
g_{\rm MPE} ( w_{\parallel} ) = Z\bar{n}_{\rm MPE} \left( \frac{m_{e}}{2\pi  T_e} \right)^{1/2} \exp \left(- \frac{m_e w_{\parallel}^2}{2T_e} \right) \Theta \left( w_{\parallel} + \sqrt{-\frac{2e \phi_{\rm W}}{m_{e}}} \right) \text{.} 
\end{align}
where $\Theta$ is the Heaviside step function,
\begin{align} \label{Heaviside}
\Theta ( \xi ) = \begin{cases}
1  \text{ for } \xi \geqslant 0 \text{,} \\
0  \text{ for } \xi < 0 \text{.} \\
\end{cases}
\end{align}
Assuming $\text{erf}(\sqrt{-e\phi_{\rm W}/T_{\rm e}}) \simeq 1$, which will be justified in the next paragraph, we obtain
\begin{align} \label{nbar}
\bar{n}_{\rm MPE} = \frac{2n_{\rm MPE} }{\left( 1+\text{erf}\left( \sqrt{- e\phi_{\rm W} / T_{\rm e} } \right) \right)} \simeq n_{\rm MPE} \text{.}
\end{align}
 
The electron current $j_{\rm e\parallel} $ is obtained from the first moment of the distribution function (\ref{g-electron}) (the flux of electrons) multiplied by the electron charge, $-e$.
The current directed towards the wall is the geometric projection of the parallel current, $j_{e, x} = - j_{\rm e\parallel} \sin \alpha \simeq - \alpha j_{e\parallel}$,
\begin{align}  \label{je-x}
j_{e, x} \simeq  \alpha Z e n_{\rm MPE} \left( \frac{T_e}{2\pi m_e} \right)^{1/2} \exp \left( \frac{e\phi_{\rm W}}{T_{\rm e}} \right)   \text{.}
\end{align}
Since the electron charge is negative and the electron flow is directed towards the wall (negative), the electron current is directed away from the wall (positive).
The electron and ion current are assumed to be similar in size.
To be consistent with the Chodura condition \citep{Chodura-1982} at the magnetic presheath entrance, the ion current is assumed to be of the order of the sound speed, giving $ j_{ e, x} \sim  \alpha Z en_{\rm MPE} c_{\rm s}$.
Hence, the electrostatic potential at the wall is 
\begin{align} \label{phiW-ordering}
\frac{e\phi_{\rm W} }{ T_{\rm e} }  \sim \ln \left( \sqrt{ \frac{2\pi m_{\rm e} (1+\tau) }{ m_{\rm i} } }  \right) 
\end{align}
where $\sqrt{2\pi m_{\rm e} (1+\tau) / m_{\rm i} } \ll 1$, justifying $\text{erf}(\sqrt{-e\phi_{\rm W}/T_{\rm e}}) \simeq 1$.

The electron distribution function at any point in the magnetic presheath and Debye sheath is \citep{Stangeby-2012}
\begin{align} \label{g-electron-x}
g ( x, w_{\parallel} ) \simeq Z n_{\rm MPE} \left( \frac{m_{e}}{2\pi  T_e} \right)^{1/2} \exp \left(\frac{e \phi (x) }{T_{\rm e}} - \frac{ m_e w_{\parallel}^2 }{2T_e} \right)  \Theta \left(   w_{\parallel} + \sqrt{\frac{2e ( \phi (x) - \phi_{\rm W} ) }{m_{e}}}  \right) \text{.} 
\end{align}
Hence, the electron density is
\begin{align} \label{n-electron-full}
n_{\rm e} ( x ) \simeq  \frac{1}{2} \left( 1 +  \text{erf} \left( \sqrt{\frac{e ( \phi (x) - \phi_{\rm W} ) }{T_{e}}} \right) \right) Z n_{\rm MPE} \exp \left(\frac{e \phi (x) }{T_{\rm e}} \right) \text{.} 
\end{align}
In the magnetic presheath the electrostatic potential is at its smallest at the Debye sheath entrance, $\lambda_{\rm D} \ll x \ll \rho_{\rm s}$, where $\phi (x) \simeq \phi_{\rm DSE}$. 
Thus, provided $ \text{erf} \left( \sqrt{e ( \phi_{\rm DSE} - \phi_{\rm W} ) / T_{e}} \right) \simeq 1$, the electron density in the magnetic presheath is given by the Boltzmann distribution 
\begin{align} \label{n-electron-MPS}
n_{\rm e} ( x ) \simeq  Z n_{\rm MPE} \exp \left(\frac{e \phi (x) }{T_{\rm e}} \right) \text{.} 
\end{align}

We proceed to justify equation (\ref{n-electron-MPS}).
The ion flow speed parallel to the magnetic field at the magnetic presheath entrance, $\rho_{\rm s} \ll x \ll d_{\rm c}$, is of the order of the sound speed $\sim c_{\rm s}$.
Projecting this parallel flow in the direction normal to the target gives $\alpha c_{\rm s} \sim \alpha \sqrt{1+\tau} v_{\rm B}$.
The ion velocity component perpendicular to the magnetic field averages to zero at the magnetic presheath entrance, as the electric field is small and the target is too far away to capture ions during their gyromotion.
Conversely, at the Debye sheath entrance the size of the ion flow is determined by the ordering for the velocity component normal to the target, $v_x \sim \sqrt{ 1 + \alpha \tau } v_{\rm B}$.
Since the number of ions in the magnetic presheath is conserved in steady state, the ion flux into the magnetic presheath, $\alpha n_{\rm MPE} \sqrt{1+\tau} v_{\rm B} $, and the ion flux out of the magnetic presheath, $n_{\rm DSE} \sqrt{1+\alpha \tau } v_{\rm B}$, are equal.
The ion density at the Debye sheath entrance is thus $n_{\rm DSE} \sim \alpha n_{\rm MPE} \sqrt{1+\tau}  / \sqrt{1+\alpha \tau } $.
Hence, we find
\begin{align} \label{phiDSE-ordering}
\frac{ e \phi_{\rm DSE} }{ T_{e} } \sim  \ln \left( \frac{ \alpha \sqrt{1+\tau} }{  \sqrt{1+\alpha \tau } } \right) \rm 
\end{align}
and
\begin{align} \label{phiDS}
\frac{ e (  \phi_{\rm W} - \phi_{\rm DSE} ) }{ T_{e} } \sim  \ln \left( \frac{1}{\alpha} \sqrt{\frac{2\pi m_{\rm e} (1+\alpha \tau)}{m_{\rm i}} } \right) \rm .
\end{align}
%which are consistent with previous work \citep{Stangeby-2012}.
Upon neglecting the factors of $\alpha \tau$, the estimates in (\ref{phiW-ordering}), (\ref{phiDSE-ordering}) and (\ref{phiDS}) are consistent with the ones in \cite{Stangeby-2012}.
Equation (\ref{n-electron-MPS}) follows from expanding equation (\ref{n-electron-full}), with $\phi(x) \geqslant \phi_{\rm DSE}$, using the orderings (\ref{phiDS}) and $ \sqrt{Zm_{\rm e}  / m_{\rm i}}  \ll \alpha$.
Note that (\ref{phiW-ordering}), (\ref{phiDSE-ordering}) and (\ref{phiDS}) are all negative, with the arguments of the logarithm smaller than unity.

The results of this section that will be used in the rest of the paper are equation (\ref{n-electron-MPS}) for the electron density in the magnetic presheath, and equation (\ref{je-x}) for the relationship between electron current and wall potential.

\section{Ion trajectories} \label{sec-trajectories}

In this section the trajectories of ions in the magnetic presheath and the Debye sheath are analyzed in detail.
The goal of this section is to relate the velocity of an ion at the target to the energy and magnetic moment of its circular gyro-orbit at the magnetic presheath entrance $\rho_{\rm s} \ll x \ll d_{\rm c}$. 
We analyze the ion trajectories first in the magnetic presheath, section \ref{subsec-traj-MPS}, and then in the Debye sheath, section \ref{subsec-traj-DS}.

\subsection{In the magnetic presheath} \label{subsec-traj-MPS}

We proceed to focus on the magnetic presheath, where $x\sim \rho_{\rm s}$.
Ions move under the influence of a wall-normal electrostatic electric field and a magnetic field at an angle $\alpha$ with the wall.
The ion equations of motion are
 \begin{align} \label{EOM-vx-full}
 \dot{v}_x = - \frac{\Omega \phi'(x)}{B}  + \Omega v_y  \cos\alpha   \text{,}
 \end{align}
\begin{align} \label{EOM-vy-full}
 \dot{v}_y = - \Omega v_x  \cos\alpha -  \Omega v_z \sin \alpha  \text{,}
 \end{align} 
 \begin{align} \label{EOM-vz-full}
 \dot{v}_z = \Omega v_y  \sin \alpha   \text{.}
 \end{align} 
 For \emph{grazing} angles, $\alpha \ll 1$, the equations simplify to
  \begin{align} \label{EOM-vx}
 \dot{v}_x \simeq   -  \frac{\Omega \phi'(x)}{B} + \Omega v_y   \text{,}
 \end{align}
\begin{align} \label{EOM-vy}
 \dot{v}_y \simeq - \Omega v_x   -  \alpha \Omega v_z   \text{,}
 \end{align} 
 \begin{align} \label{EOM-vz}
 \dot{v}_z \simeq \alpha \Omega v_y  \text{,}
 \end{align} 
 where only small terms linear in $\alpha$ were retained.
It will be useful to introduce two \emph{orbit parameters},
\begin{align} \label{xbar-def}
\bar{x} = x + \frac{v_y}{\Omega} \text{,}
\end{align}
\begin{align} \label{Uperp-def}
U_{\perp} = \frac{1}{2} v_x^2 +  \frac{1}{2} v_y^2 + \frac{\Omega \phi(x)}{B} \text{,}
\end{align}
whose time derivatives satisfy $\dot{\bar{x}} \simeq - \alpha v_z $ and $\dot{U}_{\perp} \simeq - \alpha \Omega v_y v_z $.
The third orbit parameter,
\begin{align} \label{U-def}
U = \frac{1}{2} v_x^2  + \frac{1}{2} v_y^2  + \frac{1}{2} v_z^2  + \frac{\Omega \phi(x)}{B} \text{,}
\end{align}
is just the total energy of an ion and is exactly conserved, $\dot{U} = 0$.
From the definitions (\ref{xbar-def})-(\ref{U-def}), we obtain 
\begin{align} \label{vz-U-Uperp}
v_z = \sqrt{2\left( U - U_{\perp} \right) } \text{,}
\end{align}
\begin{align} \label{vy-xbar-x}
v_y = \Omega (\bar{x} - x)   \text{,}
\end{align}
and
\begin{align}  \label{vx-xbar-Uperp-x}
v_x = \pm \sqrt{2\left( U_{\perp} -  \chi (x, \bar{x}) \right) } \text{.}
\end{align}
In (\ref{vx-xbar-Uperp-x}) an effective potential function,
\begin{align} \label{chi}
\chi(x, \bar{x}) = \frac{1}{2} \Omega^2 \left( x - \bar{x} \right)^2 + \frac{\Omega \phi(x)}{B} \text{,}
\end{align} 
was introduced.
Note that, to lowest order in $\alpha \ll 1$, $v_z$ is equivalent to the velocity component parallel to the magnetic field.
The electric field slowly (due to the grazing angle) pushes ions in the direction parallel to the magnetic field towards larger $v_z$ \citep{Geraldini-2017}.
All ions enter the magnetic presheath with a parallel velocity directed towards the target, and so they have $v_z \geqslant 0$ to lowest order in $\alpha$.
Since the parallel velocity towards the wall increases in the magnetic presheath, ions with $v_z < 0$ are not present.
Therefore, in (\ref{vz-U-Uperp}) we have set $v_z \geqslant 0$.

The orbit parameter $\bar{x}$ is referred to as the \emph{orbit position}, and $U_{\perp}$ as the \emph{perpendicular energy} (perpendicular to the magnetic field).
Since $\dot{\bar{x}}/ \rho_{\text{s}} \sim  \dot{U}_{\perp} / c_{\text{s}}^2 \sim \alpha \Omega \ll \Omega$, the orbit position and perpendicular energy only change by a very small amount during the timescale $\sim 1/\Omega$.
Neglecting the small change in the orbit parameters (which is a good approximation for a time $\ll 1/(\alpha \Omega )$), particle orbits are solved for as follows.
Consider a stationary point of the effective potential, $\chi_{\text{st}}(\bar{x}) = \chi (x_{\text{st}}, \bar{x})$, such that $\chi'(x_{\text{st}}, \bar{x}) = \Omega^2 (x_{\rm st} - \bar{x} ) + \Omega \phi' (x_{\rm st}) / B=0$. 
Here, it is understood that $\chi'(x, \bar{x}) = \partial \chi (x, \bar{x}) / \partial x $.
Rearranging this equation gives the orbit parameter as a function of the position of a stationary point,
\begin{align} \label{chimin}
 \bar{x} = x_{\text{st}} + \frac{ \phi'(x_{\rm st})}{\Omega B} \text{.} 
\end{align}
A stationary point is a minimum, $x_{\rm st} = x_{\rm m}$, if $\chi''(x_{\text{m}}, \bar{x}) = \Omega^2 + \Omega \phi'' (x_{\rm m}) / B > 0$, leading to
\begin{align} \label{cond-min}
 \phi''(x_{\rm m}) > -\Omega B \text{.}
\end{align}
At the magnetic presheath entrance, the electrostatic potential is assumed to monotonically converge to the value $\phi_{\rm MPE} = 0$.
We further assume that $\phi''(x)$ is negative (the magnitude of the electric field, $\phi'(x)$, decreases away from the wall) and monotonically converges to zero at the magnetic presheath entrance.
Hence, the stationary point is a minimum for $x_{\rm st} > x_{\rm c}$, where $ \phi''(x_{\rm c}) = -\Omega B$ if $\phi''_{\rm DSE} \leqslant - \Omega B$ or $\lambda_{\rm D} \ll x_{\rm c} \ll \rho_{\rm s}$ if $\phi''_{\rm DSE} > - \Omega B$.
Here $\phi''_{\rm DSE} $ denotes $\phi''(x)$ at the Debye sheath entrance, $\lambda_{\rm D} \ll x \ll \rho_{\rm s}$, and $x_{\rm c}$ is a critical point corresponding to the inflection point of $\chi$, if it exists, or the Debye sheath entrance $\lambda_{\rm D} \ll x_{\rm c} \ll \rho_{\rm s}$.
There are either two or one solutions for stationary points of the effective potential according to equation (\ref{chimin}), depending on whether the function $x + \phi'(x) / (\Omega B)$ has a stationary point or not. 
This leads to the distinction between two orbit types in the magnetic presheath.
Type I orbits occur when the effective potential $\chi (x, \bar{x})$ has only one stationary point: a minimum $x_{\rm m}$.
Type II orbits occur when $\chi (x, \bar{x})$ has two stationary points: a minimum $x_{\rm m}$ and a maximum $x_{\rm M} < x_{\rm m}$.
For $\bar{x} > \phi'_{\rm DSE} / (\Omega B)$, where $\phi'_{\rm DSE} $ denotes $\phi'(x)$ at the Debye sheath entrance, there is only one solution to equation (\ref{chimin}) in the magnetic presheath and therefore there are only type I ion orbits.
For both type I and type II orbits, the motion is periodic in the neighbourhood of the minimum.
The turning points $x_{\rm b} $ (for ``bottom'') and $x_{\rm t}$ (for ``top'') of the periodic motion satisfy $x_{\rm M} \leqslant x_{\rm b} < x_{\rm m}$ and $x_{\rm t} > x_{\rm m}$.
They are obtained by solving for the positions at which $v_x = 0$, i.e. $U_{\perp} = \chi ( x_{\rm b, t} , \bar{x} )$.

The slow change in $\bar{x}$ and $U_{\perp}$ cannot be entirely neglected, as it leads to ions eventually reaching the wall.
Ion trajectories are approximately periodic over a short timescale, $\sim 1/\Omega$.
Over a long enough timescale, $\sim 1/(\alpha\Omega )$, the effect of the slow variation in $\bar{x}$ and $U_{\perp}$ becomes significant.
Nonetheless, the quasi-periodic motion of the ion has an adiabatic invariant
\begin{align} \label{mu-xbar-Uperp}
\mu = \frac{1}{\pi} \int_{x_{\rm b}}^{x_{\rm t}} \sqrt{2\left(U_{\perp} - \chi(x, \bar{x}) \right)} dx \text{,}
\end{align}
which is conserved to lowest order in $\alpha \ll 1$ during the entire ion trajectory in the magnetic presheath \citep{Cohen-Ryutov-1998, Geraldini-2017}.
At the magnetic presheath entrance, $\rho_{\rm s} \ll x \ll d_{\rm c}$, $\phi (x) = 0$ and so the adiabatic invariant of (\ref{mu-xbar-Uperp}) is given by $\mu = (1 / \pi ) \int_{x_{\rm b}}^{x_{\rm t}} ds \sqrt{2 U_{\perp} - \Omega^2 (s - \bar{x} )^2  }$ with $x_{\rm b} = \bar{x} - \sqrt{2U_{\perp} }/ \Omega $ and $x_{\rm t} = \bar{x} + \sqrt{2U_{\perp} }/ \Omega $.
Upon changing variables to $\varphi$ using $s = \bar{x} - ( \sqrt{2U_{\perp}} / \Omega ) \cos \varphi$, the adiabatic invariant becomes $\mu = \left( 2U_{\perp} / (\pi  \Omega ) \right)  \int_{0}^{\pi} d\varphi  \sin^2 \varphi = U_{\perp} / \Omega $.
Using this result and equation (\ref{Uperp-def}) for $U_{\perp}$, with $\phi (x) = 0$, we obtain $\mu = \left( v_x^2 + v_y^2 \right)/ ( 2 \Omega )$.
This is equivalent to the magnetic moment to lowest order in $\alpha \ll 1$; the small difference is geometric and arises because $v_x$ is not exactly perpendicular to the magnetic field. 

The ion motion can be described as approximately periodic only insofar as it is not about to be interrupted by the absorbing wall. 
If the perpendicular energy becomes larger than a threshold value, the ion gyro-orbit becomes sufficiently large that the bottom bounce point disappears.
The threshold value of $U_{\perp}$ is the maximum value of the effective potential function between the position of the minimum, $x=x_{\rm m}$, and the wall, $x=0$, 
\begin{align} \label{chiM-def}
\chi_{\text{M}}(\bar{x}) \equiv \chi (x_{\text{M}}, \bar{x}) = \max_{x\in[0,x_{\text{m}}]} \chi (x, \bar{x}) \text{.}
\end{align}
For type I orbits, the effective potential maximum lies at the Debye sheath entrance $\lambda_{\rm D} \ll x_{\rm M} \ll \rho_{\rm s}$, such that $\chi_{\rm M} (\bar{x}) \simeq \Omega^2 \bar{x}^2 / 2 + \Omega \phi_{\rm DSE} / B$. 
For type II orbits, the effective potential maximum lies in the magnetic presheath $x_{\rm M} \sim \rho_{\rm s}$, such that $\chi_{\rm M} (\bar{x}) = \Omega^2 (x_{\rm M}-\bar{x})^2 / 2 + \Omega \phi(x_{\rm M}) / B$. 
In this case, $x_{\rm M}$ is a stationary point.
Since the variation of $U_{\perp}$ and $\bar{x}$ is slow compared to the timescale of ion motion, ions quickly reach the wall once $U_{\perp} > \chi_{\rm M} (\bar{x})$, and therefore these ions have $U_{\perp} \simeq \chi_{\rm M} (\bar{x})$.
Any ion reaching the wall must --- since it comes from an approximately periodic orbit --- have a value of orbit position such that an effective potential minimum exists.
From equation (\ref{chimin}), the smallest value of orbit position, denoted $\bar{x}_{\rm c}$, for ions in the magnetic presheath is 
\begin{align} \label{xbarc-def}
\bar{x}_{\rm c} = \min \left( x + \frac{\phi'(x)}{\Omega B} \right) = x_{\rm c} + \frac{ \phi'(x_{\rm c})}{  \Omega B } \rm .
\end{align}
Note that the second equality defines the value of $x_{\rm c}$, which is consistent with the discussion after equation (\ref{cond-min}) where $x_{\rm c}$ is first introduced.

\subsection{In the Debye sheath} \label{subsec-traj-DS}

Here, we focus on ions in the Debye sheath, $x \sim \lambda_{\rm D} \ll \rho_{\rm s}$.
Considering $\bar{x} \sim \rho_{\rm s}$ and neglecting $x \ll  \rho_{\rm s}$ in equation (\ref{vy-xbar-x}) gives
\begin{align} \label{vy-DS}
v_y \simeq \Omega \bar{x} \rm .
\end{align}
For every ion in the Debye sheath, we can trace back its trajectory to a quasiperiodic orbit.
The associated value of $\mu$ is a function of $\bar{x} \text{ (} \simeq v_y / \Omega \text{)}$ only, since $U_{\perp} \simeq \chi_{\rm M}(\bar{x})$ for ions reaching the target,
\begin{align} \label{mu-op}
\mu_{\rm op} (\bar{x}) = \frac{1}{\pi} \int_{x_{\text{M}}}^{x_{\text{t}}} \sqrt{2\left(\chi_{\text{M}} (\bar{x}) - \chi(x, \bar{x}) \right)} dx \text{.}
\end{align}
Here we have used $x_{\rm b} = x_{\rm M}$ for $U_{\perp} = \chi_{\rm M} (\bar{x})$.
The value of $v_z$ is determined by the total energy $U$, 
\begin{align} \label{vz-DS}
v_z \simeq \sqrt{2\left( U - \chi_{\rm M} (\bar{x} )\right)} \rm .
\end{align}

In order to calculate $v_x$ in the Debye sheath, the final piece of the ion trajectory in the magnetic presheath must be considered.
This is a transition from a quasiperiodic orbit, with at least one turning point in its future trajectory, to an open orbit, with no turning points in its future trajectory.
The small change of $\bar{x}$ and $U_{\perp}$ causes the value of $U_{\perp} - \chi_{\text{M}} (\bar{x})$ to increase until $U_{\perp} > \chi_{\rm M} (\bar{x})$.
The increase is slow and so the change in $U_{\perp} - \chi_{\rm M} (\bar{x})$ incurred by an ion transitioning from $U_{\perp} < \chi_{\rm M} (\bar{x})$ to $U_{\perp} > \chi_{\rm M} (\bar{x})$ can be calculated approximately by assuming a periodic orbit with fixed $U_{\perp} = \chi_{\rm M} (\bar{x})$, as shown in Appendix \ref{app-open}.
Such an orbit is fictitious: it has a bottom turning point coinciding with the position of the effective potential maximum, $x_{\rm M}$, and for a type II orbit it takes an infinite time to turn around at $x_{\rm M}$.
The true orbit turns at $x_{\rm b} > x_{\rm M}$ (with $U_{\perp} < \chi_{\rm M}$), then once more at $x_{\rm t}$, and then passes $x_{\rm M}$ (with $U_{\perp} > \chi_{\rm M}$)  in a finite time $\sim \ln (1/\alpha ) / \Omega$ moving towards the wall.
Yet, despite the approximate orbit being qualitatively different from the true orbit, the change in $U_{\perp} - \chi_{\rm M} (\bar{x})$ is accurate to lowest order in $\alpha$ when calculated from the approximate orbit.
This is because the long time spent near $x_{\rm M}$ does not contribute to a significant change in $U_{\perp} - \chi_{\rm M} (\bar{x})$, as the time derivatives of $U_{\perp}$ and of $\chi_{\rm M} (\bar{x})$ coincide at $x=x_{\rm M}$.
The overall change in the quantity $U_{\perp} - \chi_{\rm M}(\bar{x})$ during the last gyro-orbit is 
\begin{align} \label{DeltaM-def}
\Delta_{\rm M}(\bar{x}, U) = 2\pi \alpha V_{\parallel} \left( \chi_{\text{M}}(\bar{x}), U \right) \mu_{\rm op}'(\bar{x})  \text{,}
\end{align}
where $\mu_{\rm op}'(\bar{x})= d\mu_{\rm op}(\bar{x})/d\bar{x}$.
Equation (\ref{DeltaM-def}) is derived in Appendix \ref{app-open}.
 
 The implication of this discussion for ion trajectories in the Debye sheath is that there is a band of possible values of $v_x$ for a given value of $\bar{x}$ (or $\mu$) and $U$.
Considering $v_x^2 \simeq 2U_{\perp} - \Omega^2  \bar{x}^2 - 2\Omega \phi (x) / B $, which follows from (\ref{vx-xbar-Uperp-x}), (\ref{chi}) and $x\sim \lambda_{\rm D} \ll \rho_{\rm s}$, we obtain the range
\begin{align} \label{vx-spread}
 \chi_{\rm M}(\bar{x}) - \frac{1}{2} \Omega^2 \bar{x}^2 - \frac{\Omega \phi (x) }{B}   \leqslant  \frac{v_x^2}{2} <  \chi_{\rm M}(\bar{x}) + \Delta_{\rm M}(\bar{x},U) - \frac{1}{2} \Omega^2 \bar{x}^2 - \frac{\Omega \phi (x) }{B}   \text{.} 
\end{align}
Equation (\ref{vx-spread}) is valid at any point in the Debye sheath, including the Debye sheath entrance and the target.
For $\sqrt{Zm_{\rm e} / m_{\rm i}} \ll 1$ the Debye sheath repels most electrons from the wall and attracts all ions to the wall, so ions in the Debye sheath must have $v_x < 0$.

\section{Ion velocity distribution} \label{sec-IVDF}

The ion distribution function at the magnetic presheath entrance, $\rho_{\rm s} \ll x \ll d_{\rm c}$ is denoted $f_{\rm MPE}(v_x, v_y, v_z )$. 
The exact distribution function in this region includes a small number of ions with $v_z<0$, that are travelling out of the magnetic presheath towards the collisional presheath.
However, to lowest order in $\rho_{\rm s} \ll d_{\rm c}$ there are no such ions,
\begin{align}
f_{\rm MPE} (v_z < 0) = 0 \rm .
\end{align}
It can be shown that the distribution function is independent of the gyrophase angle \citep{Cohen-Ryutov-1998, Geraldini-2017} and therefore can be expressed in the form $F(\mu, U ) $.
The relationship between $f_{\rm MPE}$ and $F$ is obtained by recalling that $\mu = \left( v_x^2 + v_y^2 \right)/ ( 2 \Omega )$ at the magnetic presheath entrance,
\begin{align}
f_{\rm MPE} (v_x, v_y, v_z ) = F\left( \frac{  v_x^2 + v_y^2 }{ 2 \Omega } , \frac{  v_x^2 + v_y^2  + v_z^2}{ 2  } \right) \rm .
\end{align}
The function $F(\mu, U ) $ is conserved across the magnetic presheath to lowest order in $\alpha \ll 1$, since $\mu$ and $U$ are conserved.

The ion density at the magnetic presheath entrance, denoted $n_{\rm MPE}$, is
\begin{align} \label{normalization}
n_{\rm MPE} = 2\pi \int_0^{\infty} \Omega d\mu \int_{\Omega \mu}^{\infty} \frac{F(\mu, U) dU}{\sqrt{2\left( U - \Omega \mu \right) }} = \int f_{\rm MPE} (v_x, v_y, v_z ) d^3 v \rm .
\end{align}
The ion current towards the wall, $j_{\text{i},x}$, is obtained from the projection of the flow in the direction parallel to the magnetic field.
For $\alpha \ll 1$, this is approximately equal to 
\begin{align} \label{current-ion-parallel}
\frac{j_{\text{i},x}}{Ze} \simeq - 2\pi \alpha \int_0^{\infty} \Omega d\mu \int_{\Omega \mu}^{\infty} F(\mu, U) dU = - \alpha \int f_{\rm MPE} (v_x, v_y, v_z ) v_z d^3 v  \rm .
\end{align}
We define the total current normal to the wall as
\begin{align} \label{current-total}
j_{x}= j_{\text{e},x} + j_{\text{i}, x} \rm .
\end{align}
From equations (\ref{je-x}) and (\ref{current-total}), the electrostatic potential at the wall is
\begin{align} \label{current-potential}
\exp \left( \frac{e\phi_{\rm W}}{T_{\rm e}} \right)  \simeq \frac{ - j_{\text{i}, x} + j_{x} }{ \alpha Z e n_{\rm MPE} }  \sqrt{ \frac{2\pi m_e}{T_e} }  \rm .
\end{align}
The ion current is determined by (\ref{current-ion-parallel}), which leads to
\begin{align} \label{phi-wall}
\frac{e\phi_{\rm W} }{ T_{\rm e} }  \simeq \ln \left[ \sqrt{\frac{2\pi m_e}{T_e}}  \left( \frac{1}{n_{\rm MPE}} 2\pi \int_0^{\infty} \Omega d\mu \int_{\Omega\mu}^{\infty} dU F \left(\mu, U\right) + \frac{j_x }{\alpha Z en_{\rm MPE}} \right) \right] \rm .
\end{align}
The numerical results of this paper, presented in section \ref{sec-numerical}, are obtained assuming ambipolarity, $j_x = 0$.

As was shown in section \ref{sec-trajectories}, every value of $\mu$ and $U$, originally associated with a circular gyro-orbit entering the magnetic presheath, is associated with a specific value of $v_y \simeq \Omega \bar{x}$ and $v_z \simeq \sqrt{2\left( U - \chi_{\rm M} (\bar{x}) \right) }$ at the Debye sheath entrance, where $\mu = \mu_{\rm op}(\bar{x})$.
Here, $v_x$ is given by equation (\ref{vx-spread}) with $\phi (x) = \phi_{\rm DSE}$.
Conservation of the phase space distribution function $F(\mu, U)$ leads to the following velocity distribution \citep{Geraldini-2018}, 
\begin{align} \label{fopen-DSE}
f_{\rm DSE} (v_x, v_y, v_z) \simeq & ~ F \left( \mu_{\rm op} ( \bar{x}) , U \right) \Theta \left( \bar{x} - \bar{x}_{\text{c}} \right)  \Theta \left( -v_x \right)   \nonumber \\ & \times \hat{\Pi} \left(  \frac{1}{2}v_x^2 - \chi_{\rm M} ( \bar{x} ) + \frac{1}{2} \Omega^2 \bar{x}^2 + \frac{\Omega \phi_{\rm DSE}}{B}  ,  0 , \Delta_{\rm M}(\bar{x}, U )  \right) \text{.}
\end{align}
Here, we have defined the top-hat function 
\begin{align}
\hat{\Pi} ( \xi, \xi_1, \xi_2 ) = \begin{cases} 1 \text{ for } \xi_1 \leqslant \xi < \xi_2 \text{, } \\
0 \text{ else.} \end{cases} 
\end{align}
In Appendix \ref{app-ionconservation} it is shown that the ion current normal to the wall calculated from (\ref{fopen-DSE}) is equal to (\ref{current-ion-parallel}), and thus (\ref{fopen-DSE}) satisfies ion conservation.
At the wall, where $x=0$, the range of possible values of $v_x$ associated with each value of $\bar{x}$ and $U$ is given by equation (\ref{vx-spread}) with $\phi (0) = \phi_{\rm W}$,
\begin{align} \label{fopen-W}
f_{\rm W} (v_x, v_y, v_z) \simeq & ~ F \left( \mu_{\rm op} ( \bar{x}) , U \right) \Theta \left( \bar{x} - \bar{x}_{\text{c}} \right)  \Theta \left( -v_x \right)   \nonumber \\ & \times \hat{\Pi} \left(  \frac{1}{2}v_x^2 - \chi_{\rm M} ( \bar{x} ) + \frac{1}{2} \Omega^2 \bar{x}^2 + \frac{\Omega \phi_{\rm W}}{B}  ,  0 , \Delta_{\rm M}(\bar{x}, U )  \right) \text{.}
\end{align}

In order to obtain $f_{\rm DSE}$, and consequently $f_{\rm W}$, it is necessary to determine the constants $\bar{x}_{\rm c}$ and $\phi_{\rm DSE}$, and the functions $\chi_{\rm M}(\bar{x})$ and $\mu_{\rm op}(\bar{x})$.
Recall that, by equation (\ref{DeltaM-def}), $\chi_{\rm M}(\bar{x})$ and $\mu_{\rm op}(\bar{x})$ also determine $\Delta_{\rm M}( \bar{x}, U )$.
These quantities are specified by the electrostatic potential profile $\phi (x)$, which is obtained by solving the quasineutrality equation (\ref{quasineutrality}).
Thus, equation (\ref{fopen-DSE}) does not --- \emph{per se} --- fully specify $f_{\rm DSE} (v_x, v_y, v_z) $.
In \cite{Geraldini-2018} an expression for the ion density $n_{\rm i} (x)$ for $\alpha \ll 1$, as a functional of the electrostatic potential $\phi (x)$, was derived.
Using this expression, an iterative scheme to obtain the numerical solution $\phi (x)$ of the quasineutrality equation (\ref{quasineutrality}) was presented.
In the next section, a model for $f_{\rm DSE} (v_x, v_y, v_z) $ is presented, which allows to bypass obtaining a numerical solution of $\phi (x)$ across the whole magnetic presheath.

\section{Large ion gyro-orbit model} \label{sec-VDMP-model}

In this section we derive a closed set of equations for the quantities $\bar{x}_{\rm c}$, $\phi_{\rm DSE}$, $\chi_{\rm M}(\bar{x})$ and $\mu_{\rm op}(\bar{x})$ appearing in equations (\ref{fopen-DSE}) for $f_{\rm DSE}$ and (\ref{fopen-W}) for $f_{\rm W}$.
The derivation assumes $ \tau \gg 1 $ and exploits the approximately undistorted nature of ion gyro-orbits in this limit. 
In section~\ref{subsec-VDMP-quasi}, the quasineutrality equation is expanded in the magnetic presheath close to the Debye sheath entrance, $\lambda_{\rm D} \ll x \ll \rho_{\rm s}$, to obtain a relationship between the distribution function and electric field. 
Then, in section~\ref{subsec-VDMP-largetraj} the expression for the electric field is used to derive expressions for the functions $\chi_{\rm M}(\bar{x})$ and $\mu_{\rm op}(\bar{x})$.
This procedure is strictly not self-consistent, as the expression for the electric field derived in the previous subsection is valid closer to the wall than where it is used.
To determine the large gyro-orbit distribution function, only the two parameters $\bar{x}_{\rm c}$ and $\phi_{\rm DSE}$ remain to be specified. 
In section \ref{subsec-VDMP-closure}, a method to solve for the two parameters is presented. 

\subsection{Quasineutrality at the Debye sheath entrance} \label{subsec-VDMP-quasi}

In general, solving equation (\ref{quasineutrality}) in the magnetic presheath is a numerical task. 
However, near the Debye sheath entrance the quasineutrality equation can be expanded to obtain analytical expressions relating the electric field to the distribution function in this region.
This analysis is valid for $\sqrt{Zm_{\rm e} / m_{\rm i} } \ll \alpha$, as it assumes equation (\ref{n-electron-MPS}) for the electron density.

The variation in density in the magnetic presheath, close to the Debye sheath entrance, for both ions and electrons is related to the variation in the electrostatic potential, $\delta \phi (x) = \phi (x) - \phi_{\rm DSE} $.
The Boltzmann distribution (\ref{n-electron-MPS}) is expanded near the Debye sheath entrance to obtain 
\begin{align}
n_{\rm e} (x) \simeq  Z n_{\rm MPE} \exp \left( \frac{e \phi_{\rm DSE} }{T_{\rm e}} \right) \left( 1 +  \frac{e\delta \phi}{T_{\rm e}} + \left( \frac{e\delta \phi}{T_{\rm e}} \right)^2 \right)  \rm .
\end{align}
The form of the expansion of the ion density in $\delta \phi (x)$ depends on whether ions with $v_x=0$ are present or not at the Debye sheath entrance, i.e. whether $f_{\rm DSE} (v_x= 0) = 0$ or not.
If $f_{\rm DSE} (v_x= 0) \neq 0$, equation (\ref{fopen-DSE}) requires that $\chi_{\rm M} (\bar{x}) = \Omega^2 \bar{x}^2 / 2 + \Omega \phi_{\rm DSE} / B$ for at least some values of $\bar{x}$, i.e. type I ion orbits must be present.
Thus, there are ions whose bottom turning point lies very close to the Debye sheath entrance at $x_{\rm b} \leqslant x$.
Such ions have a velocity range between $|v_x| = 0$ ($x_{\rm b} = x$) and $|v_x| = \sqrt{2\left( \chi_{\rm M } (\bar{x}) - \chi (x, \bar{x} ) \right)} \simeq \sqrt{2\left( \Omega^2 \bar{x} x - \Omega \delta \phi (x) / B \right)} \sim \sqrt{\delta \phi}$ ($x_{\rm b} \simeq 0$), and can have both positive and negative values of $v_x$.
%Conversely, ions at the Debye sheath entrance must have $v_x \leqslant 0$.
%Therefore, the additional ions with a bottom bounce point 
These ions contribute to a term in the ion density proportional to $\sqrt{\delta \phi}$ \citep{Geraldini-2018}, heuristically due to the size of the additional integration region in $v_x$.
Since no term in the electron density is proportional to $\sqrt{\delta \phi}$, type I ion orbits must be absent, requiring 
\begin{align} \label{bc-nobackDS}
f_{\rm DSE} (v_x= 0) = 0 \rm .
\end{align}
Recall from section \ref{sec-trajectories} that all ions with $\bar{x} > \phi'_{\rm DSE} / (\Omega B )$ --- corresponding to a sufficiently large value of $\mu = \mu_{\rm op} ( \bar{x})$ --- are in type~I orbits.
For there to be a complete absence of type I orbits, $\phi'(x)$ must be divergent at the Debye sheath entrance on the magnetic presheath scale, $  \phi'_{\rm DSE}  \rightarrow \infty $.\footnote{The divergence in $\phi'(x)$ is resolved by retaining the term $\epsilon_{0} \phi''(x)$, small in $\lambda_{\rm D} / \rho_{\rm s} \ll 1$, in Poisson's equation (\ref{Poisson}).}
As shown in the next subsection, this divergence also causes the asymptotic distribution function $f_{\rm DSE}$ to decay exponentially for $v_x \rightarrow 0$ provided $F (\mu, U)$ decays exponentially for $U \rightarrow \infty$.

Excluding the presence of type I orbits, the ion density near the Debye sheath entrance is obtained by following ion characteristics backwards from the Debye sheath entrance.  
To lowest order in $\alpha$, the orbit parameters $\bar{x}$ and $U_{\perp}$ are constant; in addition, the total energy $U$ is exactly constant.
Consider equations (\ref{vz-U-Uperp}), (\ref{vy-xbar-x}) and (\ref{vx-xbar-Uperp-x}) for the ion velocity in the magnetic presheath. 
The quantities $v_z $, $v_y + \Omega x $ and $-\sqrt{v_x^2 + 2\Omega \delta \phi (x) / B - 2\Omega^2 \bar{x} x + \Omega^2 x^2 } $ are constant and, from equations (\ref{vy-DS}), (\ref{vz-DS}) and (\ref{vx-spread}), are equal to the components of the velocity at the Debye sheath entrance. 
Thus, the ion density at a distance $x$ from the wall, near the Debye sheath entrance, is
\begin{align} \label{ni-DSE-1}
n_{\rm i} (x) \simeq \int f_{\rm DSE}\left( -\sqrt{v_x^2 + \frac{2\Omega \delta \phi (x) }{ B } - 2\Omega v_y x  }, v_y + \Omega x, v_z \right)  d^3 v \rm .
\end{align}
Here, we have neglected the term $ \Omega^2 x^2 \ll 2\Omega v_y x$.
The quasineutrality equation (\ref{quasineutrality}) to lowest order in $e\delta \phi (x) /T_{\rm e} \ll v_x^2 / v_{\rm B}^2$ and $x \ll v_x^2 / (\Omega v_y ) \sim v_x^2 / (\Omega c_{\rm s} )$\footnote{For $v_y = \Omega \bar{x} \gg c_{\rm s}$ the distribution function is exponentially small provided it is exponentially decaying at large energies, and therefore the typical value $v_y \sim c_{\rm s}$ can be used.} gives an equation for $\phi_{\rm DSE}$, 
\begin{align} \label{quasi-DSE}
 n_{\rm DSE}  \equiv \int f_{\rm DSE} ( \vec{v} ) d^3 v  = n_{\rm MPE} \exp \left( \frac{e \phi_{\rm DSE} }{T_{\rm e}} \right) \rm .
\end{align}
In (\ref{quasi-DSE}) we have denoted the lowest-order ion density at the Debye sheath entrance as $n_{\rm DSE}$.

Considering the exponential decay of $f_{\rm DSE}$ for $|v_x | \rightarrow 0$, the first argument of $f_{\rm DSE} $ in (\ref{ni-DSE-1}) can be expanded in $e\delta \phi (x ) / T_{\rm e} \ll v_x^2 / v_{\rm B}^2$ and $x \ll v_x^2 / (\Omega c_{\rm s} )$ to give
\begin{align} \label{ni-DSE-2}
n_{\rm i} (x) \simeq \int f_{\rm DSE}\left( v_x + \frac{\Omega \delta \phi }{ B v_x } - \frac{ \Omega v_y x }{ v_x } - \frac{\Omega^2 \delta \phi^2  }{ 2B^2 v_x^3 }  , v_y + \Omega x, v_z \right)  d^3 v \rm .
\end{align}
The result of Taylor expanding the integrand in (\ref{ni-DSE-2}) and subsequently integrating by parts is
\begin{align}
n_{\rm i} (x) \simeq  \int f_{\rm DSE} ( \vec{v} ) d^3 v + \frac{\Omega \delta \phi}{ B }  \int \frac{  f_{\rm DSE} ( \vec{v} ) }{ v_x^2 }  d^3v - \Omega x  \int \frac{ v_y f_{\rm DSE} ( \vec{v} ) }{ v_x^2 }  d^3v \nonumber \\ + \frac{3}{2} \left( \frac{ \Omega \delta \phi }{ B } \right)^2 \int \frac{f_{\rm DSE} ( \vec{v} )  }{ v_x^4 }  d^3 v    \rm .
\end{align}
An alternative derivation of the same result is obtained by integrating the top-hat function in $v_x$ first and then expanding the resulting expression \citep{Geraldini-2018}.
Note that the Taylor expansion of the second argument of $f_{\rm DSE}$ in equation (\ref{ni-DSE-1}), $v_y + \Omega x$, about $v_y$ did not give a variation in $x$.
Collecting terms that are higher order than (\ref{quasi-DSE}) in the quasineutrality equation gives an equation relating electrostatic potential variation and position,
\begin{align} \label{quasi-exp}
~~& \frac{e\delta \phi}{T_{\rm e}} \left(  \int f_{\rm DSE} ( \vec{v} ) d^3 v  - v_{\rm B}^2 \int \frac{f_{\rm DSE} ( \vec{v} ) }{ v_x^2 }  d^3v  \right) \nonumber  \\
+ & \frac{1}{2} \left( \frac{e\delta \phi}{T_{\rm e}} \right)^2 \left(  \int f_{\rm DSE} ( \vec{v} ) d^3 v  - 3 v_{\rm B}^4 \int \frac{f_{\rm DSE} ( \vec{v} )  }{ v_x^4 }  d^3 v \right)  +  \Omega x  \int \frac{f_{\rm DSE} ( \vec{v} ) v_y }{ v_x^2 }  d^3v  \simeq 0    \rm .
\end{align}
Since, as was concluded in the previous paragraph, the electric field must diverge for $x \rightarrow 0$, the appropriate balance of terms in equation (\ref{quasi-exp}) is $\delta \phi^2 \propto x$.
Therefore, the term linear in $\delta \phi$ must be set to zero, and we obtain the marginal form of the kinetic Bohm condition \citep{Geraldini-2018},
\begin{align} \label{Bohm}
I_{\rm Bohm} \equiv v_{\rm B}^2 \int \frac{f_{\rm DSE} ( \vec{v} )}{v_x^2} d^3 v  =  n_{\rm DSE}  \text{.}
\end{align} 
In (\ref{Bohm}) we have defined the Bohm integral, $I_{\rm Bohm}$, and we have used the definition of $ n_{\rm DSE} $ in (\ref{quasi-DSE}).

The condition (\ref{Bohm}) applies to the \emph{lowest-order} distribution function in the region $\lambda_{\rm D} \ll x \ll  \rho_{\rm s} $.
It does not apply to the exact distribution function measured near a target in an experiment \citep{Riemann-2012-comment, Baalrud-Hegna-2012-reply}.
There are small corrections to the asymptotic distribution function $f_{\rm DSE} ( \vec{v} )$ in the region $\lambda_{\rm D} \ll x \ll \rho_{\rm s}$. 
With a finite but large electric field, $\phi'(x)$, the distribution function in this region does not exactly satisfy $f(x, \vec{v})= 0$ for $v_x = 0$. 
One reason for this is the presence of a small number of very high-energy ions whose bottom turning point is only a few Debye lengths from the target, $x_{\rm b} \sim \lambda_{\rm D}$.
A very small number of ion collisions or reflections from the target, both neglected, would also cause $f(x,\vec{v}) \neq 0$ for $v_x \geqslant 0$.
If the exact distribution function, $f (x, \vec{v} )$, were used instead of the asymptotic one, $f_{\rm DSE} (\vec{v})$, in the kinetic Bohm condition (\ref{Bohm}), then the left hand side would diverge, $ \int ( f (x, \vec{v} ) / v_x^2 ) d^3 v \rightarrow \infty \rm $, and the condition could not even be approximately satisfied.
Nonetheless, $f_{\rm DSE} (\vec{v})$ is --- within the validity of the underlying orderings --- an approximation of the true distribution function in the region $\lambda_{\rm D} \ll x \ll \rho_{\rm s}$.

Imposing (\ref{Bohm}), equation (\ref{quasi-exp}) becomes
\begin{align}
 \left( \frac{e\delta \phi}{T_{\rm e}} \right)^2 \left( \int f_{\rm DSE} ( \vec{v} ) d^3 v  - 3 v_{\rm B}^4 \int \frac{f_{\rm DSE} ( \vec{v} )  }{ v_x^4 }  d^3 v \right) +  2 \Omega x  \int \frac{f_{\rm DSE} ( \vec{v} ) v_y }{ v_x^2 }  d^3v  \simeq 0    \rm .
\end{align}
The electrostatic potential variation in the magnetic presheath, near the Debye sheath entrance, is thus given by 
\begin{align} \label{phi-near0}
\frac{e\left( \phi ( x ) - \phi ( 0 ) \right) }{T_{\rm e}}  \simeq \frac{ \sqrt{2 \bar{x}_{\rm av} x} }{\rho_{\rm B}} \text{,}
\end{align}
with $\bar{x}_{\rm av}$, denoting a kinetic average of $\bar{x} = v_y / \Omega$, given by
\begin{align} \label{xbarav-def}
\frac{  \bar{x}_{\rm av} }{\rho_{\rm B}} = \frac{  v_{\rm B}  \int \left( v_y f_{\text{DSE}} \left( \vec{v}  \right) / v_x^2  \right) d^3 v  }{   \int f_{\text{DSE}} \left( \vec{v}  \right) \left( 3  v_{\text{B}}^4 / v_x^4 - 1 \right) d^3 v }   \sim \frac{\sqrt{1+\tau}}{(1+ \alpha \tau ) }   \text{.}
\end{align}
Here, $\rho_{\rm B} = v_{\rm B} / \Omega$ is referred to as the Bohm gyroradius.
Since $f_{\rm DSE}$ is exponentially small near $v_x=0$, the integral in the denominator of (\ref{xbarav-def}) is convergent. 
The ordering in (\ref{xbarav-def}) can be obtained as follows.
Consider the smallest value of $ |v_x |$ in the range (\ref{vx-spread}) at the Debye sheath entrance ($\phi (x) = \phi_{\rm DSE}$),
\begin{align} \label{Vxslow}
V_{x,\rm slow} (\bar{x}) = \sqrt{2\left( \chi_{\rm M} (\bar{x}) - \frac{1}{2} \Omega^2 \bar{x}^2 - \frac{\Omega \phi_{\rm DSE}}{B} \right) } \rm .
\end{align}
Ions with $v_x  \simeq - V_{x, \rm slow}$ are referred to as ``slow'' ions.
From equation (\ref{quasi-DSE}) and the ordering $|v_x| \sim v_{\rm B} \sqrt{1+\alpha \tau}$ for typical values of $v_x$, the marginalized distribution function is ordered $\int f_{\rm DSE} dv_y dv_z \sim n_{\rm DSE} / ( v_{\rm B}\sqrt{1 + \alpha \tau} )$.
The kinetic Bohm condition (\ref{Bohm}) determines the size of slow ions, $\int (f_{\rm DSE} / v_x^2 ) d^3 v \sim \left( \int f_{\rm DSE} dv_y dv_z \right) / V_{x,\rm slow} \sim  n_{\rm DSE} / (V_{x,\rm slow} v_{\rm B}  \sqrt{1+\alpha \tau} )$. 
This gives the ordering $V_{x,\rm slow} \sim v_{\rm B} / \sqrt{1+\alpha \tau}$.
Note that $V_{x,\rm slow} \ll  v_{\rm B} \sqrt{1+\alpha \tau}$ only if $\alpha \tau \gg 1$, so that for $\alpha \tau \lesssim 1$ the normal velocity of slow ions is similar in size to the normal velocity of a typical ion.
The size of $\bar{x}_{\rm av}$ is obtained by considering the contribution of slow ions to the integrals in (\ref{xbarav-def}) and using also $v_y \sim c_{\rm s}$, giving $\bar{x}_{\rm av} / \rho_{\rm B} \sim c_{\rm s} V_{x,\rm slow}^2 / v_{\rm B}^3 \sim \sqrt{1+\tau} / (1+\alpha \tau) $.

The region of validity of equation (\ref{phi-near0}) is obtained by investigating the validity of the expansion (\ref{quasi-exp}).
In order for the expansion to be valid, the orderings $e\delta \phi (x ) / T_{\rm e} \ll v_x^2 / v_{\rm B}^2$ and $x \ll v_x^2 / (\Omega c_{\rm s} )$ must be satisfied.
Using $x \ll V_{x,\rm slow}^2 / (\Omega c_{\rm s} )$, the ordering $x \ll \rho_{\rm s} / [(1+\tau ) ( 1 + \alpha \tau ) ]$ for the region of validity of the expansion is obtained.
The same ordering results from $e\delta \phi (x ) / T_{\rm e} \ll V_{x,\rm slow}^2 / v_{\rm B}^2$ using equations (\ref{phi-near0}) and (\ref{xbarav-def}).

\subsection{Ion trajectories and ion distribution function for $ \tau \gg 1$} \label{subsec-VDMP-largetraj}
 
In order to obtain $f_{\rm DSE} (\vec{v})$ from (\ref{fopen-DSE}), the electrostatic potential in the magnetic presheath is necessary to calculate: the function $\chi_{\rm M} (\bar{x})$ from equation (\ref{chiM-def}), the function $\mu_{\rm op}(\bar{x})$ from equation (\ref{mu-op}), the quantity $\bar{x}_{\rm c}$ from equation (\ref{xbarc-def}) and the quantity $\phi_{\rm DSE}$.
These quantities are calculated here using a model obtained by considering ion trajectories for $\tau \gg 1$ in the electrostatic potential of equation (\ref{phi-near0}).

For $ \tau \gg 1$, the thermal velocity of an ion is much larger than the Bohm velocity, $v_{\rm t,i}^2 \sim \tau v_{\rm B}^2  \gg v_{\rm B}^2 $.
To calculate the adiabatic invariant, we can therefore neglect the small electrostatic potential variation throughout the orbit, $\Omega \phi (x) / B \sim v_{\rm B}^2 \ll \Omega \mu_{\rm op}(\bar{x}) \sim v_{\rm t,i}^2 $, and using equation (\ref{mu-op}) obtain $\mu_{\rm op} (\bar{x}) \simeq U_{\perp} / \Omega \simeq \chi_{\rm M} ( \bar{x} ) / \Omega$.
This does not specify the functional form of $\mu_{\rm op} (\bar{x})$ and $\chi_{\rm M} ( \bar{x} )$, but in relating them reduces the number of unknown functions from two to one. 
The approximate equivalence of $U_{\perp}$ and $\Omega \mu$ and the conservation of $U$ and $\mu$ imply that $v_z = \sqrt{2\left(U - U_{\perp} \right)}$, has remained approximately unchanged from its value at the magnetic presheath entrance, $\sqrt{2\left(U - \Omega \mu \right)}$.
The quantity $\bar{x}_{\rm c}$, defined in (\ref{xbarc-def}), corresponds to the orbit position of a gyro-orbit with adiabatic invariant equal to zero (since $x_{\rm b} = x_{\rm t} = x_{\rm c}$), and thus $\bar{x}_{\rm c}$ is obtained through $\mu_{\rm op} (\bar{x}_{\rm c}) =0$. 

When an ion in a large gyro-orbit gets sufficiently close to the target, its gyro-motion is distorted as shown in figure \ref{fig-ion}(b).
The net force away from the wall on an ion at a given instant is given by the effective potential gradient, $\chi'(x, \bar{x})$.
The distortion of ion gyro-orbits is caused by a competition between the magnetic force pulling away from the wall and the electric force pushing towards the wall.
Since type I orbits are absent, $x_{\rm M} $ is a stationary point where the electric force on the ion exactly balances the magnetic force.
Its location can be obtained from equation (\ref{chimin}) with $x_{\rm st} = x_{\rm M} < x_{\rm c}$, 
\begin{align} \label{chiMax}
\bar{x} =  x_{\rm M} + \frac{ \phi'(x_{\rm M})}{\Omega B}  \rm .
\end{align}

In what follows, the electrostatic potential in (\ref{phi-near0}) is used to approximate the electrostatic potential at distances from the wall corresponding to typical values of $x_{\rm M}$.
From (\ref{phi-near0}) we obtain $\phi'(x_{\rm M} ) = (T_{\rm e} / e \rho_{\rm B} ) \sqrt{\bar{x}_{\rm av} / 2x_{\rm M} }$.
Using the ordering $x_{\rm M} \ll \bar{x} \sim \phi'(x_{\rm M})/\Omega B \sim \rho_{\rm s}$ in equation (\ref{chiMax}), we obtain
\begin{align} \label{xM-xbar-largemu}
x_{\text{M}} = \frac{\bar{x}_{\rm av} \rho_{\text{B}}^2}{2 \bar{x}^2} \rm{.}
\end{align}
By inserting (\ref{xM-xbar-largemu}) into $\chi_{\rm M} ( \bar{x}) = \Omega^2 (x_{\rm M} -\bar{x} )^2 / 2 + \Omega \phi (x_{\rm M}) / B $, neglecting the term $\Omega^2 x_{\rm M}^2 / 2$ and remembering that $\mu_{\rm op} (\bar{x}) \simeq \chi_{\rm M} ( \bar{x} ) / \Omega$, we obtain
\begin{align} \label{mu-chiM-large}
\Omega \mu_{\rm op} (\bar{x} ) \simeq \chi_{\rm M}(\bar{x})  \simeq  \frac{1}{2} \Omega^2 \bar{x}^2 + \frac{v_{\rm B}^2  \bar{x}_{\rm av}}{2\bar{x}} +  \frac{\Omega \phi_{\rm DSE} }{ B }   \text{.} 
\end{align}
Imposing $\mu_{\rm op} (\bar{x}_{\rm c} ) = 0$ in equation (\ref{mu-chiM-large}) gives 
\begin{align} \label{xbarc-model}
\bar{x}_{\rm c} =   \rho_{\rm B} \sqrt{-\frac{2e \phi_{\rm DSE}}{T_{\rm e}} -  \frac{v_{\rm c}^2}{v_{\rm B}^2} }   \text{,}
\end{align}
where the quantity $v_{\rm c} = v_{\rm B}  \sqrt{ \bar{x}_{\rm av} / \bar{x}_{\rm c}} $, called the critical velocity, has been defined.
With this definition, $\bar{x}_{\rm av}$ is given by 
\begin{align} \label{xbarav-model}
\bar{x}_{\rm av} =  \frac{ v_{\rm c}^2 }{v_{\rm B} ^2} \bar{x}_{\rm c}  \rm .
\end{align}

From equations (\ref{mu-chiM-large}) and (\ref{vx-spread}), large gyro-orbits at the Debye sheath entrance have a range of normal velocities given by
\begin{align} \label{vx-spread-large}
 \frac{v_{\rm c}^2  \bar{x}_{\rm c}}{2\bar{x}}  \leqslant   \frac{v_x^2}{2} <   \frac{v_{\rm c}^2  \bar{x}_{\rm c}}{2\bar{x}} + 2 \pi \alpha  \mu_{\rm op }'(\bar{x}) \sqrt{2\left( U - \Omega \mu_{\rm op}(\bar{x}) \right)}  \text{.} 
\end{align}
Inserting the velocity spread (\ref{vx-spread-large}) in the distribution function (\ref{fopen-DSE}) the velocity distribution of ions in large gyro-orbits is
\begin{align} \label{f-large-DSE}
f_{\rm DSE} (v_x, v_y, v_z) \simeq & ~ F \left(  \mu_{\rm op}(\bar{x})  , U \right) \Theta \left( \bar{x} - \bar{x}_{\text{c}} \right)  \Theta \left( -v_x \right)   \nonumber \\ & \times \hat{\Pi} \left(  \frac{1}{2}v_x^2 -  \frac{v_{\rm c}^2  \bar{x}_{\rm c}}{2\bar{x}} , ~ 0 ,  ~2 \pi \alpha  \mu_{\rm op }'(\bar{x}) \sqrt{2\left( U - \Omega  \mu_{\rm op}(\bar{x}) \right)}    \right) \text{.}
\end{align}

Despite being a useful analytical model for the ion distribution function, the large gyro-orbit model presented here is strictly not asymptotically self-consistent.
For $\tau \gg 1$, $\bar{x} \sim \rho_{\rm i}$ and $\bar{x}_{\rm av} \sim \rho_{\rm i} / (1+\alpha \tau )$, where $\rho_{\rm i} = v_{\rm t,i}/ \Omega$ is the thermal ion gyroradius.
Using equation (\ref{xM-xbar-largemu}), it follows that $x_{\rm M}  \sim \rho_{\rm i} /\left[  \tau ( 1+\alpha \tau ) \right]$.
Recall from the final paragraph of section \ref{subsec-VDMP-quasi} that the expansion used to derive equation (\ref{phi-near0}) is valid, for $\tau \gg 1$, in the region $x_{\rm M} \ll  \rho_{\rm i} /\left[  \tau ( 1+\alpha \tau ) \right]$.
Therefore, equation (\ref{xM-xbar-largemu}) is not valid for the majority of ions.
There is, however, a minority of ions for which $\mu \gg v_{\rm t,i}^2 / \Omega$ and $\bar{x} \gg \rho_{\rm i}$, which have $x_{\rm M}  \ll  \rho_{\rm i} /\left[  \tau ( 1+\alpha \tau ) \right]$.
For these ions equation (\ref{xM-xbar-largemu}) is accurate. 
This can be used to derive the exponential decay of $f_{\rm DSE}$ at $|v_x| \rightarrow 0$ as follows.
The distribution function $F (\mu, U )$ is assumed to exponentially decay for $U \rightarrow \infty$ and consequently, since $U \geqslant \Omega \mu$, for $\Omega \mu \rightarrow \infty$, such that $F \sim \exp \left( - 2\Omega \mu / v_{\rm t,i}^2 \right)  $.
If follows from (\ref{mu-chiM-large}) and $\bar{x} \rightarrow \infty$ that $F \sim \exp \left( - \Omega^2 \bar{x}^2 / v_{\rm t,i}^2 \right)$.
The slowest value of $|v_x|$ in the top-hat function in (\ref{fopen-DSE}) is given by the function $V_{x,\rm slow}$ in (\ref{Vxslow}), which in the model is
\begin{align}
V_{x,\rm slow} (\bar{x}) = v_{\rm c} \sqrt{\frac{ \bar{x}_{\rm c} }{\bar{x}}} \rm .
\end{align}
For the top-hat function in (\ref{fopen-DSE}) to be non-zero we require  $V_{x,\rm slow} (\bar{x}) \leqslant |v_x|$ and so $\bar{x} \geqslant \bar{x}_{\rm c} v_{\rm c}^2 / v_x^2$.
Therefore, the largest value of $f_{\rm DSE}$ for an ion with $v_x \rightarrow 0$ satisfies $f_{\rm DSE} \sim \exp \left( - \Omega^2 \bar{x}_{\rm c}^2 v_{\rm c}^4 /(v_{\rm t,i}^2 v_x^4 ) \right) $, which is exponentially small.

The critical velocity is the value of $| v_x |$ for an ion at the Debye sheath entrance with $\mu = 0$, which came from an infinitesimally small gyro-orbit, $v_{\rm c} = V_{x, \rm slow} (\bar{x}_{\rm c})  $.
These ions should have $\mu_{\rm op}' ( \bar{x}_{\rm c} ) = 0$ and thus $\Delta_{\rm M} (\bar{x}_{\rm c}, U ) = 0$ for all values of $U$, which would give $v_x = - v_{\rm c}$ as the only allowed value according to the velocity distribution (\ref{fopen-DSE}).
However, ions with $\mu = 0$ in the model have a finite range of velocities due to the fact that $\mu_{\rm op}' (\bar{x}_{\rm c}) = 0$ is not imposed in order not to overconstrain the model. 
This could be concerning, since if $\mu_{\rm op}'(\bar{x}_{\rm c} ) < 0 $ (and so $\Delta_{\rm M} (\bar{x}_{\rm c}, U ) < 0$) the range of values of $v_x^2$ in (\ref{vx-spread}) would allow for non-real values of $v_x$. 
Fortunately, $\mu_{\rm op}'(\bar{x}_{\rm c} ) = \Omega \bar{x}_{\rm c} - v_{\rm c}^2 / (2\Omega \bar{x}_{\rm c} )  $ is always positive if $\alpha$ is sufficiently small that $e|\phi_{\rm DSE}| / T_{\rm e} \sim \ln \alpha \gg 1$, as equation (\ref{xbarc-model}) leads to $2 (\Omega \bar{x}_{\rm c})^2 \sim |\ln \alpha | v_{\rm B}^2 \gg v_{\rm c}^2 \sim v_{\rm B}^2$.
In practice, $\mu_{\rm op}'(\bar{x}_{\rm c} ) > 0$ for all values of $\alpha \leqslant 5^{\circ}$ considered in this paper.
With $\mu_{\rm op}'(\bar{x}_{\rm c} ) > 0 $, ions with $\mu = 0$ ($\bar{x} = \bar{x}_{\rm c}$) have a non-zero range of values of $v_x$ according to equations (\ref{DeltaM-def}) and (\ref{vx-spread}).
In the model, $|v_x | = v_{\rm c}$ is therefore the smallest value of $|v_x|$ for an ion with $\mu = 0$.
Although $\mu_{\rm op}' (\bar{x}_{\rm c}) \neq 0$ may look like a serious shortcoming of the model, for $\tau \gg 1$ the large discrepancy in the function $\mu_{\rm op}' (\bar{x})$ is expected only for a small number of particles near $\bar{x} = \bar{x}_{\rm c}$. 
In other words, the model does not correctly capture the small gyro-orbits, but there are assumed to be only a small number of them anyway\footnote{The asymptotic theory of the ion trajectories is also inaccurate for small gyro-orbits, albeit not as evidently. This inaccuracy is unimportant if $\tau$ is sufficiently large that the asymptotic theory correctly describes the majority of ion orbits. It was shown in \cite{Geraldini-2019} that when $\tau \lesssim \alpha^{1/3}$ the asymptotic theory fails for an appreciable fraction of the ions.}.

\subsection{Model closure: calculating $\phi_{\rm DSE}$ and $v_{\rm c}$} \label{subsec-VDMP-closure}

The only unknowns that specify the model distribution function (\ref{f-large-DSE}) are the two constants $\phi_{\rm DSE}$ and $v_{\rm c}$. 
The value of $\phi_{\rm DSE}$ is determined from quasineutrality at the Debye sheath entrance, equation (\ref{quasi-DSE}).
The value of $v_{\rm c}$ is determined by imposing the kinetic Bohm condition (\ref{Bohm}).

For numerical evaluation, it is best to re-express all velocity moments as 
\begin{align} 
 \int f_{\rm DSE}(\vec{v}) v_x^a d^3 v = & \int_{\bar{x}_{\rm c}}^{\infty} \Omega d\bar{x} \int_{\Omega \mu_{\rm op}(\bar{x})}^{\infty} F \left( \mu_{\rm op}(\bar{x}), \Omega \mu_{\rm op}(\bar{x}) + \frac{1}{2} v_z^2  \right)  \nonumber \\ & \times  \frac{v_{\rm c}^{a+1}}{a+1}   \left(  \left(  \frac{  \bar{x}_{\rm c} }{ \bar{x} } + \frac{ 4 \pi \alpha \mu_{\rm op}'(\bar{x}) v_z }{v_{\rm c}^2} \right)^{(a+1)/2}  - \left( \frac{  \bar{x}_{\rm c} }{ \bar{x} } \right)^{(a+1)/2}   \right)   d v_z \nonumber  \text{,}
\end{align} 
obtained from (\ref{f-large-DSE}) using the change of variables $v_y = \Omega \bar{x}$ and $v_z = \sqrt{2\left( U - \Omega \mu_{\rm op}(\bar{x}) \right)}$, and substituting (\ref{xbarav-model}).
In particular, to solve equations (\ref{quasi-DSE}) and (\ref{Bohm}) for $\phi_{\rm DSE}$ and $v_{\rm c}$, we require the density,
\begin{align} \label{nDSE-model}
n_{\rm DSE} = & \int_{\bar{x}_{\rm c}}^{\infty} \Omega d\bar{x} \int_{\Omega \mu_{\rm op}(\bar{x})}^{\infty} F \left( \mu_{\rm op}(\bar{x}), \Omega \mu_{\rm op}(\bar{x}) + \frac{1}{2} v_z^2  \right)  \nonumber \\ & \times v_{\rm c}  \left(  \left(  \frac{  \bar{x}_{\rm c} }{ \bar{x} } + \frac{ 4 \pi \alpha \mu_{\rm op}'(\bar{x}) v_z }{v_{\rm c}^2} \right)^{1/2}   -  \left( \frac{  \bar{x}_{\rm c} }{ \bar{x} } \right)^{1/2}  \right)   d v_z \rm ,
\end{align}
and the Bohm integral,
\begin{align} \label{IBohm-model}
I_{\rm Bohm}  = & v_{\rm B}^2 \int_{\bar{x}_{\rm c}}^{\infty} \Omega d\bar{x} \int_{\Omega \mu_{\rm op}(\bar{x})}^{\infty} F \left( \mu_{\rm op}(\bar{x}), \Omega \mu_{\rm op}(\bar{x}) + \frac{1}{2} v_z^2  \right)  \nonumber \\ & \times  \frac{1}{v_{\rm c}}   \left( \left( \frac{  \bar{x}_{\rm c} }{ \bar{x} } \right)^{-1/2} -\left(  \frac{  \bar{x}_{\rm c} }{ \bar{x} } + \frac{ 4 \pi \alpha \mu_{\rm op}'(\bar{x}) v_z }{v_{\rm c}^2} \right)^{-1/2}  \right)   d v_z \rm .
\end{align}
Note that the value of $I_{\rm Bohm}$ decreases by increasing $v_{\rm c}$, and vice versa.

Iterative expressions are used to determine $\phi_{\text{DSE}}$ from equation (\ref{quasi-DSE}) and $v_{\rm c}$ from equation (\ref{Bohm}).
The first guesses, or zeroth iterates, are defined by $\phi_{\text{DSE},0} = (T_{\rm e} / e )\ln \alpha$ and $v_{\text{c},0} = v_{\rm B}$, and iteration values are denoted by $\phi_{\text{DSE},\nu} $ and $v_{\text{c},\nu} $.
At each iteration, $n_{\text{DSE},\nu}$ and  $I_{\text{Bohm},\nu} $ are evaluated from equations (\ref{nDSE-model}) and (\ref{IBohm-model}).
The iterates $\phi_{\text{DSE},\nu+1} $ and $v_{\text{c},\nu+1} $ are obtained using
\begin{align} \label{phiDSE-iteration}
\phi_{\text{DSE},\nu+1} = \frac{T_{\rm e} }{e} \ln \left( \frac{ n_{\text{DSE},\nu} }{ n_{\rm MPE} } \right) \rm ,
\end{align}
\begin{align} \label{vc-iteration}
v_{\text{c},\nu+1} =  & ~ \frac{v_{\text{c},\nu}}{ I_{\text{Bohm},\nu} } \left( I_{\text{Bohm},\nu} - n_{\text{DSE},\nu} \right) &  \text{ if } I_{\text{Bohm},\nu} >  n_{\text{DSE},\nu} \text{,}  \nonumber  \\
 = & ~ \epsilon_{v_{\rm c}}  & \text{ else.} 
\end{align}
Equation (\ref{phiDSE-iteration}) originates from the rearranged form of equation (\ref{quasi-DSE}), $ \phi_{\text{DSE}} = (T_{\rm e}  / e ) \ln \left(n_{\text{DSE}} / n_{\rm MPE} \right) $ .
Equation (\ref{vc-iteration}) is based on a Newton method with the approximations $d n_{\rm DSE} / dv_{\rm c} \approx 0 $ and $d I_{\rm Bohm} / dv_{\rm c}  \approx  - I_{\rm Bohm} / v_{\rm c}  $. 
The iteration is truncated when 
\begin{align} \label{cond-1}
\frac{n_{\text{DSE},N}  -  n_{\text{DSE},N-1} }{ n_{\text{DSE},N} } < \epsilon_{n} \rm ,
\end{align}
\begin{align} \label{cond-2}
\left| \frac{ I_{\text{Bohm},N}  }{ n_{\text{DSE},N} } - 1 \right | < \epsilon_{I} \rm .
\end{align}
In the earliest iterations, it may happen that $v_{\text{c},\nu +1}  \leqslant 0$, which is prevented by setting $v_{\text{c},\nu +1} $ to be a small number above zero (smaller than the solution $v_{\text{c}}$), denoted $\epsilon_{v_{\rm c}}$. 
The $N$th iteration values of $\phi_{\text{DSE}} $ and $v_{\rm c}$, satisfying conditions (\ref{cond-1}) and (\ref{cond-2}), are considered to be acceptable numerical solutions of (\ref{quasi-DSE}) and (\ref{Bohm}).
The value of $\bar{x}_{\rm av}$ is obtained from $v_{\rm c}$ using equation (\ref{xbarav-model}).
To obtain the results presented in the next section, $\epsilon_{n} = \epsilon_{I} = \epsilon_{v_{\rm c}} = 10^{-10} $ was used. 

Having solved equations (\ref{quasi-DSE}) and (\ref{Bohm}) for $\phi_{\rm DSE}$ and $v_{\rm c} = v_{\rm B}  \sqrt{ \bar{x}_{\rm av} / \bar{x}_{\rm c}} $, equations (\ref{vy-DS}), (\ref{vz-DS}), (\ref{mu-chiM-large}), (\ref{xbarc-model}) and (\ref{f-large-DSE}) completely specify the large gyro-orbit model distribution function at the Debye sheath entrance, $f_{\rm DSE}(\vec{v})$.
The model distribution function at the wall is obtained by replacing equation (\ref{f-large-DSE}) with
\begin{align} \label{f-large-W}
f_{\rm W} (\vec{v}) \simeq & ~ F \left( \mu_{\rm op}(\bar{x})  , U \right) \Theta \left( \bar{x} - \bar{x}_{\text{c}} \right)  \Theta \left( -v_x \right)   \nonumber \\ & \times \hat{\Pi} \left(  \frac{1}{2}v_x^2 - \frac{v_{\rm c}^2  \bar{x}_{\rm c}}{2\bar{x}}-  \frac{\Omega}{B} \left( \phi_{\rm DSE} - \phi_{\rm W} \right) , ~ 0 ,  ~2 \pi \alpha \Omega \bar{x} \sqrt{2\left( U - \Omega \mu_{\rm op}(\bar{x}) \right)}    \right) \text{,}
\end{align}
where equation (\ref{phi-wall}) determines the wall potential $ \phi_{\rm W}$.

To conclude this section, the application of the model to $\tau \lesssim 1$ is discussed.
We have seen that the model is derived assuming $\tau \gg 1$, although it is not asymptotically self-consistent even in this limit.
The Bohm condition closure (\ref{Bohm}) used in the model to obtain $v_{\rm c}$ (and $\bar{x}_{\rm c}$) is nonetheless valid for all $\tau$.
Therefore, for $\tau \ll 1$ the model correctly recovers a distribution function that is centred around $v_x \simeq - v_{\rm B}$, as expected from the fluid cold-ion result \citep{Chodura-1982}.
This extends the applicability of the model to smaller values of $\tau$, though with less accurate results.
A measure of the accuracy of the model can be obtained by calculating tha value of $\bar{x}_{\rm av}$ from equation (\ref{xbarav-def}) and comparing it to the model value in (\ref{xbarav-model}).
For $\alpha \tau \gg 1$, the two values are found to approach each other.
For $\tau \ll 1$ the two values are found to differ approximately (with an $O(\alpha)$ error) by a factor of two: indeed, equation (\ref{xbarav-def}) results in $\bar{x}_{\rm av} \simeq \bar{x}_{\rm c} / 2$ upon using a cold-ion distribution function centred at $v_x = - v_{\rm B}$ and $v_y = \Omega \bar{x}_{\rm c}$, while the model value from (\ref{xbarav-model}) is $\bar{x}_{\rm av} \simeq \bar{x}_{\rm c} $.

\section{Numerical results} \label{sec-numerical}

In this section, a comparison is presented of ion velocity distributions obtained from:
\begin{enumerate}
\item equations (\ref{fopen-DSE}), (\ref{fopen-W}) and the full numerical solution $\phi (x)$ of the quasineutrality equation in the magnetic presheath entrance;
\item equations (\ref{f-large-DSE}), (\ref{f-large-W}) and the closure equations of the large gyro-orbit model.
\end{enumerate} 
To obtain the solutions (i), the numerical scheme in \cite{Geraldini-2018} is used.
In section \ref{subsec-bc} the boundary conditions for the distribution function at the magnetic presheath entrance, as a function of $\tau$, are given.
Then, in section \ref{subsec-narrowing}, results for the distribution of the component $v_x$ of the ion velocity at the Debye sheath entrance, obtained using (i) and (ii), are presented.
Finally, results for the energy-angle distributions of ions at the wall are presented in section \ref{subsec-energy-angle} for some values of $\alpha$ and $\tau$.
The possibility to extend the model for $\alpha \sim \sqrt{Zm_{\rm e} / m_{\rm i}}$ is briefly discussed in section \ref{subsec-VDMP-nonadiabatic}. 

\subsection{Boundary conditions at the magnetic presheath entrance} \label{subsec-bc}

The ion velocity distribution at the magnetic presheath entrance, $\rho_{\rm s} \ll x \ll d_{\rm c}$, is taken to be
\begin{align} \label{f-infty}
f_{\rm MPE} \left( \vec{v} \right) =  \begin{cases}
\mathcal{N}  n_{\rm MPE} \frac{4 v_z^2}{\pi^{3/2} v_{\text{t,i}}^5}   \exp \left( - \frac{ \left| \vec{v} - u v_{\text{t,i}} \hat{\vec{e}}_z \right|^2 }{v_{\text{t,i}}^2} \right) \Theta \left( v_z \right) & \text{ for } \tau \leqslant 1 \text{,} \\
\mathcal{N}  n_{\rm MPE}  \frac{ 4 v_z^2 }{  \pi^{3/2} v_{\text{t,i}}^3 \left( v_{\text{t,i}}^2 +r v_z^2 \right)} \exp \left( - \frac{\left| \vec{v} \right|^2 }{v_{\text{t,i}}^2} \right)\Theta \left( v_z \right)  & \text{ for } \tau > 1 \text{,}
\end{cases}
\end{align}
for any prescribed value of $\tau$, where $\Theta $ is the Heaviside step function defined in (\ref{Heaviside}) and $\hat{\vec{e}}_z$ is a unit vector in the $z$ direction.
The family of velocity distributions (\ref{f-infty}) is the same used in \cite{Geraldini-2019} to study the dependence of the magnetic presheath solution on ion temperature, and is chosen to satisfy the marginal kinetic Chodura condition \citep{Geraldini-2018}
\begin{align} \label{kinetic-Chodura}
 v_{\text{B}}^2 \int \frac{ f_{\rm MPE} \left( \vec{v} \right)}{v_z^2} d^3v = n_{\rm MPE}  \text{.}
\end{align}
The value of the normalization constant $\mathcal{N}$ is obtained from (\ref{normalization}), giving
\begin{align} \label{N-infty}
\mathcal{N}  =  \begin{cases}
\left[  \left( 1 + 2u^2 \right) \left( 1 + \text{erf}(u) \right) + \frac{2u}{\sqrt{\pi}}  \exp(-u^2) \right]^{-1}  & \text{ for } \tau \leqslant 1 \text{,} \\
r^{3/2} \left[ 2\sqrt{r} - 2\sqrt{\pi} \exp\left(\frac{1}{r}\right) \left( 1 - \text{erf} \left( \frac{1}{\sqrt{r}} \right) \right) \right]^{-1}  & \text{ for } \tau > 1 \text{.}
\end{cases}
\end{align}
The values of $u$ and $r$ are obtained by imposing (\ref{kinetic-Chodura}), leading to
\begin{align} \label{u-def}
 1 + \text{erf}(u)  = \tau \left[ \left( 1 + 2u^2 \right) \left( 1 + \text{erf} (u) \right) + \frac{2u}{\sqrt{\pi}} \exp(-u) \right]  \text{,}
\end{align}
\begin{align} \label{r-def}
r \sqrt{\pi} \exp\left(\frac{1}{r}\right) \left( 1 - \text{erf} \left( \frac{1}{\sqrt{r}} \right) \right)  = \tau \left[  2\sqrt{r} - 2\sqrt{\pi} \exp\left(\frac{1}{r}\right) \left( 1 - \text{erf} \left( \frac{1}{\sqrt{r}} \right) \right) \right] \text{.}
\end{align}

\subsection{Narrowing of the wall-normal velocity distributions} \label{subsec-narrowing}

\begin{figure} 
\centering
\includegraphics[width=1.0\textwidth]{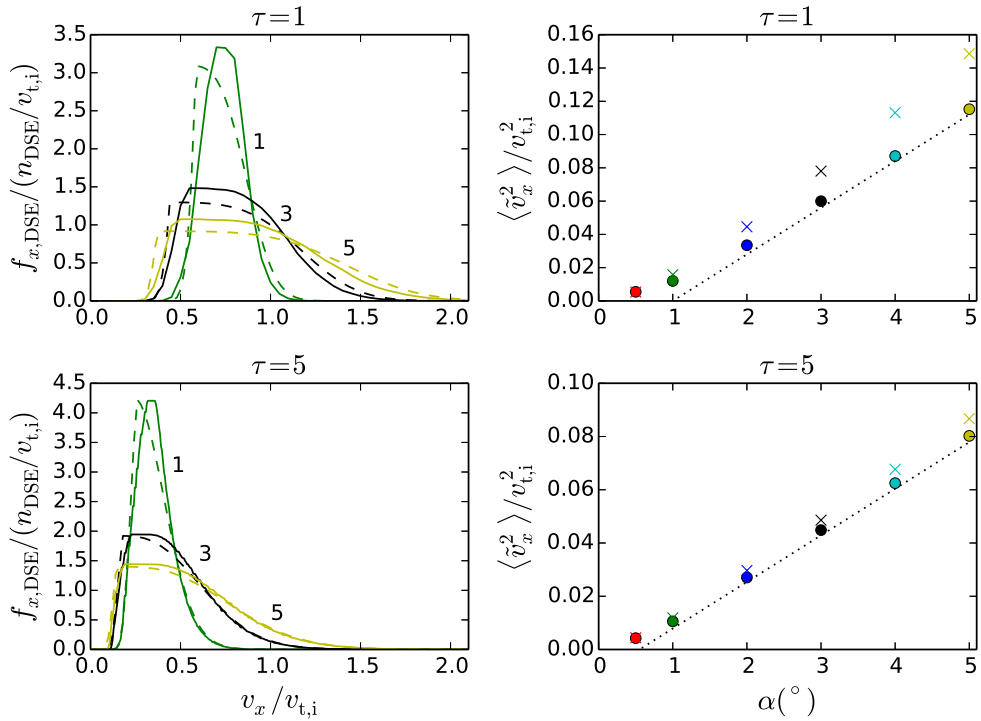}
\caption{Left: wall-normal velocity distributions at the Debye sheath entrance from the numerical solution of $\phi (x)$ in the magnetic presheath (solid lines) and from the large gyro-orbit model (dashed lines), for $\tau = 1$ (top) and $\tau = 5$ (bottom) for angles $\alpha = 1^{\circ},~3^{\circ},~5^{\circ}$.
Right: the variance $\langle \tilde{v}_x^2 \rangle $ of the distributions from the numerical solution of $\phi (x)$ (circles) and from the model (crosses) for values of $\alpha$ between $0.5^{\circ}$ and $5^{\circ}$.
The dotted lines are drawn to guide the eye, showing the linear scaling $\langle \tilde{v}_x^2 \rangle / v_{\rm t,i}^2 \sim \alpha$ for $\alpha \gtrsim 1^{\circ}$. (Note: here $\alpha$ is measured in degrees.)}
\label{fig-thinf}
\end{figure}

The marginalized distribution function
\begin{align} \label{fx}
f_{x,\rm DSE} (v_x) = \int \int f_{\rm DSE} (\vec{v}) dv_y dv_z  \rm ,
\end{align}
is the distribution of wall-normal velocities $v_x$ of ions at the Debye sheath entrance.
The numerical results obtained for $f_{x,\rm DSE} (v_x)$ with the model and the theory for $\tau = 1$ and $\tau = 5$, for a number of angles $\alpha$, are shown in figure~\ref{fig-thinf}.
The first thing to note is that the model distribution function (dashed lines) captures the essential features of the distribution function obtained from the full solution of the magnetic presheath electrostatic potential $\phi (x)$ (solid lines).
Moreover, the agreement is better for the largest value of $\tau=T_{\rm i} / ( ZT_{\rm e} )$, $\tau = 5$, as expected.

The width of the function $f_{x,\rm DSE} (v_x)$ narrows as $\alpha$ decreases, a feature that was observed in \cite{Geraldini-2018}.
The width of this function can be quantified using the variance $\langle \tilde{v}_x^2 \rangle$, defined using the second moment of $f_{x,\rm DSE} ( v_x )$,
\begin{align} \label{width-f}
\langle \tilde{v}_x^2 \rangle  = \sqrt{ \frac{\int (v_x - u_{x,\rm DSE})^2 f_{x,\rm DSE}(v_x) dv_x }{\int f_{x,\rm DSE}(v_x) dv_x } } \text{.}
\end{align}
Here
\begin{align}
u_{x,\rm DSE} = \frac{ \int v_x f_{x,\rm DSE}(v_x) dv_x }{ {\int f_{x,\rm DSE}(v_x) dv_x } }
\end{align}
is the average wall-normal velocity at the Debye sheath entrance.
As can be seen in figure~\ref{fig-thinf}, the variance of the distribution function scales linearly with $\alpha$. 

\begin{figure} 
\centering
\includegraphics[width=1.0\textwidth]{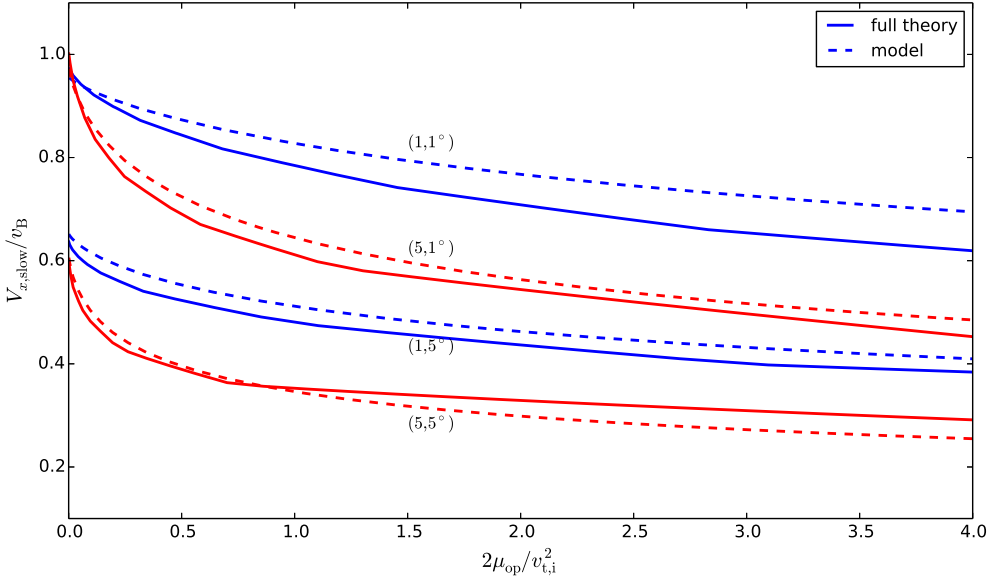}
\caption{The velocity of slow ions, $V_{x, \rm slow} (\bar{x})$, is shown as a function of the adiabatic invariant $\mu_{\rm op} (\bar{x})$ with ($\tau, \alpha $) labelled. Solid lines are obtained from the numerical solution of $\phi (x)$; dashed lines are obtained from the large gyro-orbit model. }
\label{fig-Vxslow}
\end{figure}

\begin{figure} 
\centering
\includegraphics[width=1.0\textwidth]{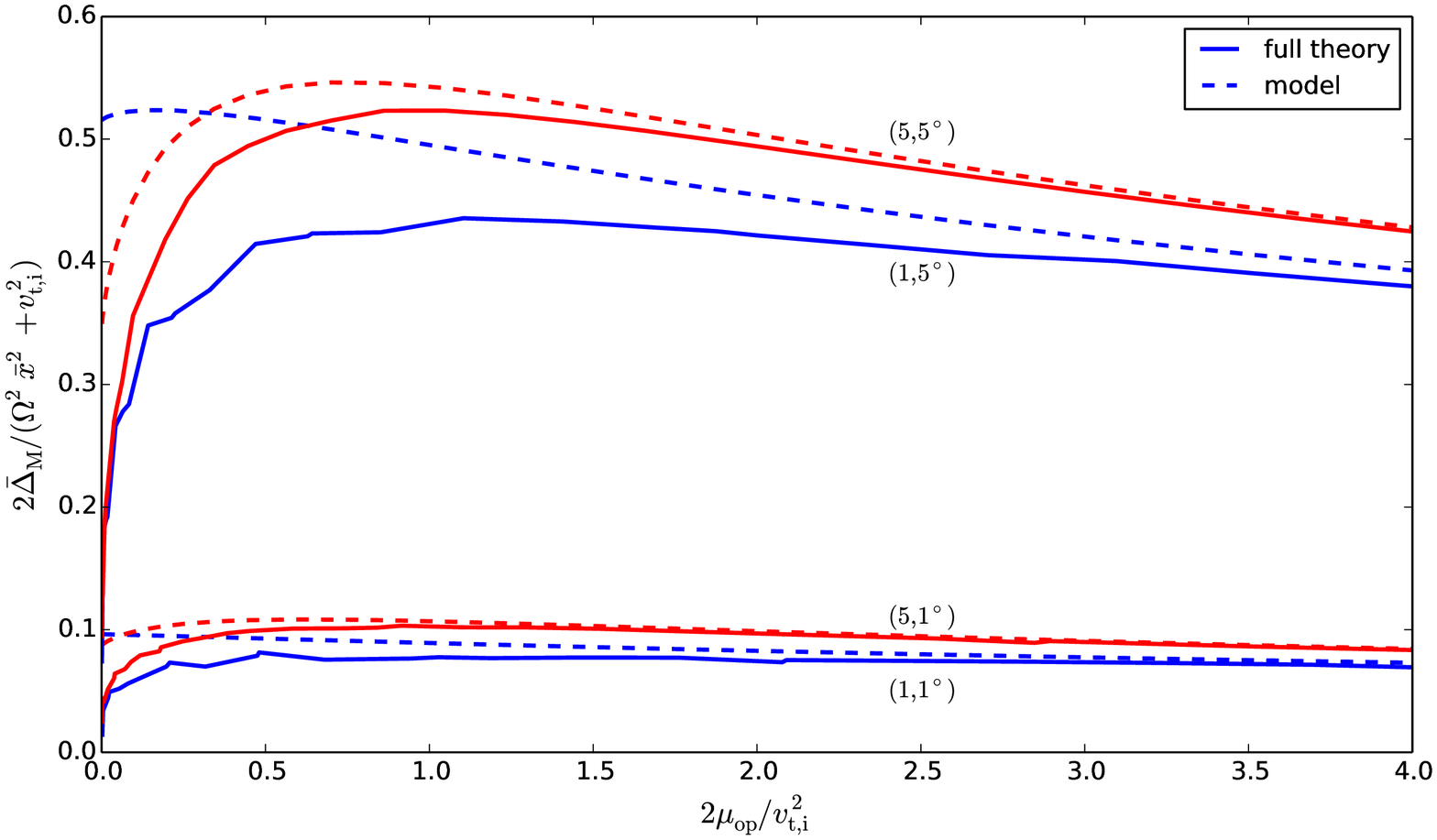}
\caption{The quantity $2 \bar{\Delta}_{\rm M} / (\Omega^2 \bar{x}^2 + v_{\rm t,i}^2 )$, with $\bar{\Delta}_{\rm M} (\bar{x}) = 2\pi \alpha \mu_{\rm op}' (\bar{x} ) v_{\rm t,i}$, is shown as a function of the adiabatic invariant $\mu_{\rm op} (\bar{x})$ for labelled values of ($\tau, \alpha$). Solid lines are obtained from the numerical solution of $\phi (x)$; dashed lines are obtained from the large gyro-orbit model. 
For $2 \bar{\Delta}_{\rm M} / (\Omega^2 \bar{x}^2 + v_{\rm t,i}^2 ) \ll 1$ the asymptotic theory in $\alpha \ll 1$ is valid.
}
\label{fig-funcs}
\end{figure}

The scaling of the variance can be explained as follows.
The ion velocity can be decomposed into two pieces: a piece coming from the electric field acceleration which depends only on $\bar{x}$ (or $\mu$), $V_{x,\rm slow} (\bar{x} ) = \sqrt{ 2\left( \chi_{\rm M}(\bar{x}) - \Omega^2 \bar{x}^2 / 2 - \Omega \phi_{\rm DSE} / B \right) } $, and an additional gyrophase dependent piece which gives the velocity range in (\ref{vx-spread}).
In figure~\ref{fig-Vxslow} the behaviour of $V_{x,\rm slow} (\bar{x} ) $ as a function of $\mu_{\rm op} (\bar{x})$ is shown for some values of $\tau$ and $\alpha$. % cases.
The slow decay of $V_{x,\rm slow} $ with $\mu$ is approximately captured by the model for $\tau = 5$, and for $(\tau, \alpha ) = ( 1, 5^{\circ} )$.
For $(\tau, \alpha ) = (1, 1^{\circ} )$, the dependence of $V_{x,\rm slow} $ on $\mu$ is stronger than predicted by the model, but is nonetheless fairly weak.
Since $V_{x,\rm slow} $ is only a weakly decreasing function of $\mu$, the distribution function sharply drops to zero around $|v_x | \approx V_{x,\rm slow} ( \rho_{\rm s} )$, a feature common to all velocity distributions in figure \ref{fig-thinf}.
The dominant contribution to the variance $\langle \tilde{v}_x^2 \rangle $ therefore comes from the range of allowed values of $|v_x |$ in equation (\ref{vx-spread}), instead of the dependence of $V_{x, \rm slow}$ on $\mu$.
For $\tau \gtrsim 1$, we order $\Omega \mu_{\rm op} \sim \Omega^2 \bar{x}^2 / 2 \sim v_{\rm t,i}^2$ and $\sqrt{2\left( U - \Omega \mu_{\rm op}(\bar{x}) \right)} \sim v_{\rm t,i}$, and obtain $2 \pi \alpha \mu_{\rm op}'(\bar{x}) \sqrt{2\left( U - \Omega \mu_{\rm op}(\bar{x}) \right)} \sim 2\pi \alpha v_{\rm t,i}^2$.
Hence, the variance is $\langle \tilde{v}_x^2 \rangle \sim \alpha v_{\rm t,i}^2 \sim \alpha \tau v_{\rm B}^2$, as seen in the numerical results.
The dependence of $V_{x,\rm slow}$ on $\mu$ does not cause a significant contribution to $\langle \tilde{v}_x^2 \rangle$ unless $\alpha \tau $ is extremely small, seen in the numerical results of figure \ref{fig-thinf} as a saturation of the decrease of the variance for $\alpha  \lesssim 1^{\circ}$.

When deriving the scaling of equation (\ref{xbarav-def}), the typical value of $| v_x |$ of slow ions was found to be $V_{x,\rm slow}  \sim v_{\rm B} / \sqrt{1+\alpha \tau }$.
From figure~\ref{fig-Vxslow} it appears that the ordering $\alpha \tau \gtrsim 1$ is satisfied, as $V_{x,\rm slow}(\bar{x})$ is smaller than $v_{\rm B}$ in most cases shown here.
It may appear concerning that $V_{x,\rm slow} / v_{\rm B}$ is quite small also for $(\tau, \alpha ) = ( 1,  5^{\circ} )$, as this suggests that $\alpha \tau $ is large for $\tau = 1$ and for a value of $\alpha$ ($=5^{\circ} \approx 0.09~ \rm radians$) which is considered small.
This observation 
prompts a closer analysis of the validity of the asymptotic theory of the ion orbits, which assumes $\alpha \ll 1$.
One of the consequences of this ordering is that the function $\Delta_{\rm M}(\bar{x}, U )$ is small.
For $\tau \gtrsim 1$, the smallness of $\Delta_{\rm M}(\bar{x}, U )$ is measured relative to the kinetic energy of the ion\footnote{For $\tau \ll 1$, enlarging ion gyro-orbits make this analysis insufficient \citep{Geraldini-2019}.}, estimated from the tangential components of the ion velocity, $(v_y^2 + v_z^2 ) / 2 \sim ( \Omega^2 \bar{x}^2 + v_{\rm t,i}^2 ) / 2$.
The ratio $2\bar{ \Delta }_{\rm M} / ( \Omega^2 \bar{x}^2 + v_{\rm t,i}^2 ) $ is shown in figure~\ref{fig-funcs} and highlights that, although for $\alpha = 5^{\circ}$ the validity of the asymptotic theory is not robust, the contribution of $\Delta_{\rm M}$ to the ion energy is smaller than the total kinetic energy for most ions, albeit by a factor of $\sim 2$ only. 
Note that $\alpha = 5^{\circ}$ corresponds to $2\pi \alpha \approx 0.6 \rm ~radians$, and so the factor of $2\pi$ in equation (\ref{DeltaM-def}) explains why the expansion in $\alpha$ starts to becomes inaccurate at $\alpha \approx 5^{\circ}$.

Although this subsection presented ion distribution functions at the Debye sheath entrance, $f_{x,\rm DSE} (\vec{v})$, the validity of the scaling $\langle \tilde{v}_x^2 \rangle \sim \alpha v_{\rm t,i}^2$ is expected to apply also to the ion velocity distribution at the wall, $f_{x,\rm W} (\vec{v})$.
In the next subsection, ion velocity distributions at the wall are considered for parameters where $\phi_{\rm DSE} > \phi_{\rm W}$, such that the assumption of Boltzmann electrons (recall equation (\ref{phiDS})) remains at least approximately correct.

\subsection{Energy-angle distributions at the target} \label{subsec-energy-angle}

Since sputtering predictions depend on the distribution of kinetic energy and angle of impact of ions reaching the target, it is useful to calculate the energy-angle distribution of ions at the wall. 
To obtain our results, we considered a Deuterium plasma such that $\sqrt{ m_{\rm e} / m_{\rm i}} = 0.0165$ and $Z=1$.

\begin{figure} 
\centering
\includegraphics[width=1.0\textwidth]{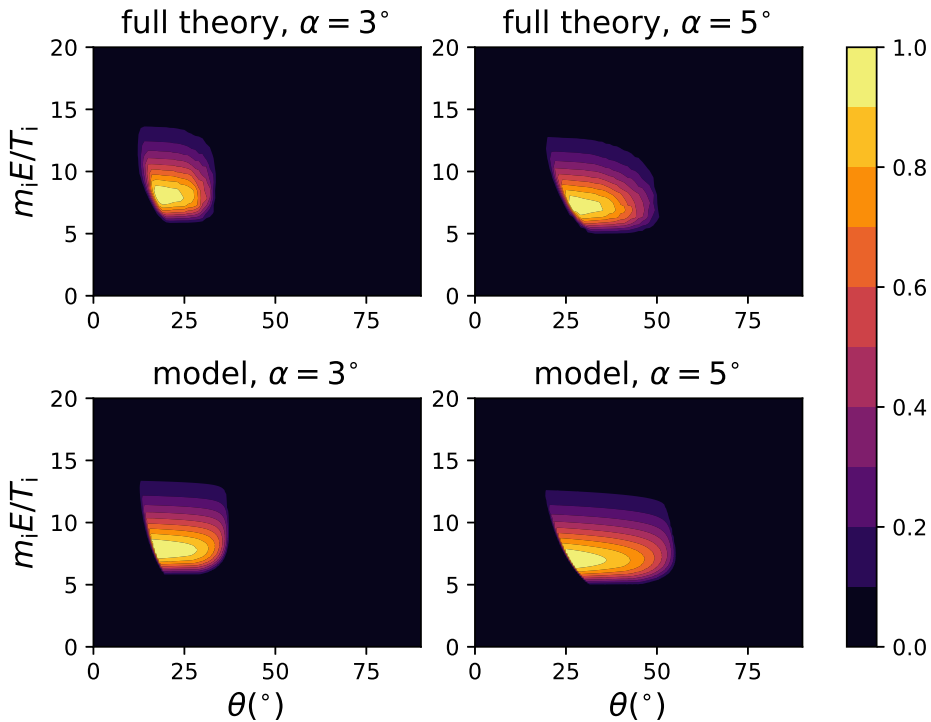}
\caption{Energy-angle distributions at the target, $\zeta_{\rm W} ( E, \theta )$, obtained from the full electrostatic potential solution (full theory) and from the large gyro-orbit model for $\tau = 0.5$ and $\alpha = 3^{\circ}$ and $5^{\circ}$, are shown normalized to their peak value.
}
\label{fig-energyangle-tau0pt5}
\end{figure}

\begin{figure} 
\centering
\includegraphics[width=1.0\textwidth]{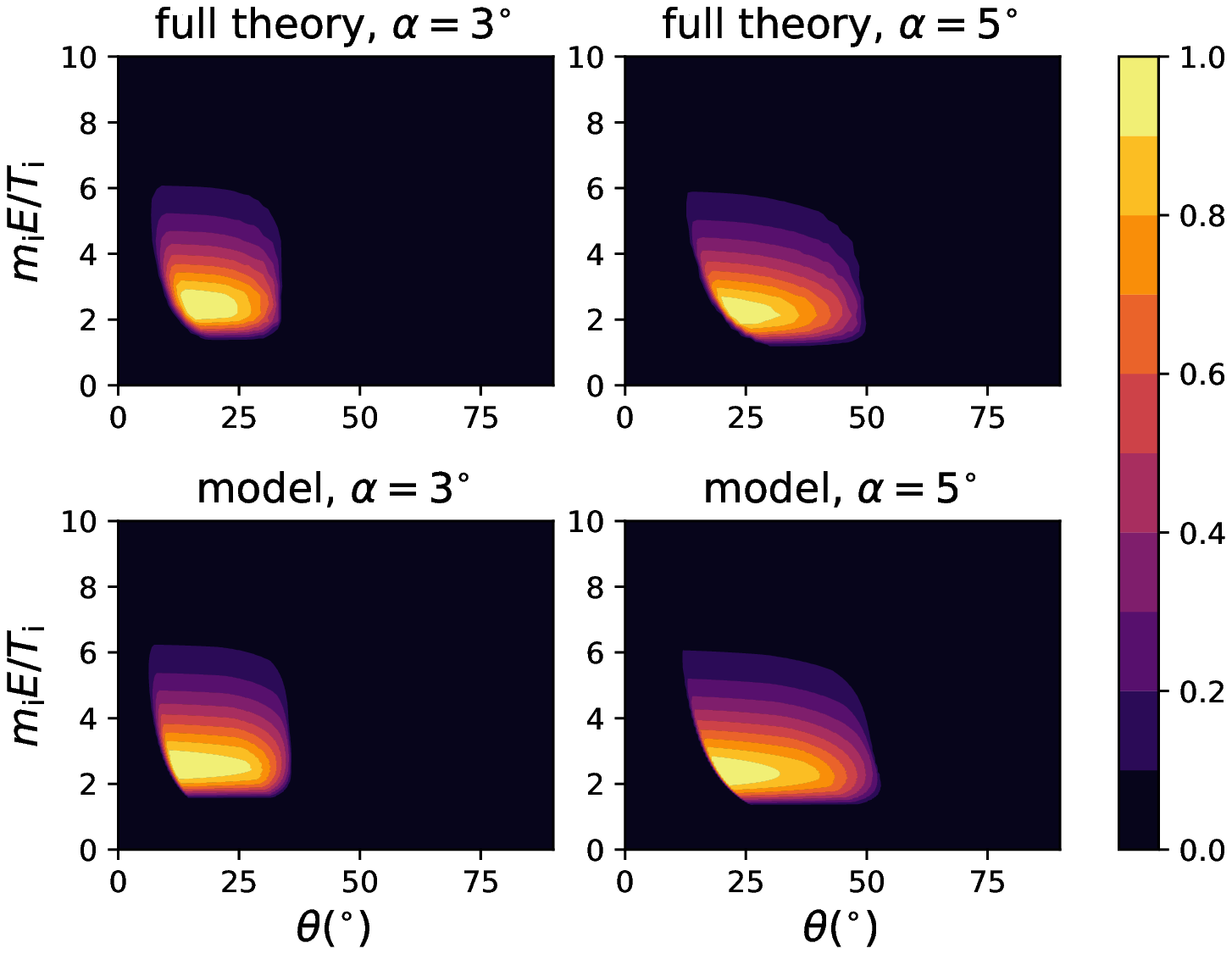}
\caption{Energy-angle distributions at the target, $\zeta_{\rm W} ( E, \theta )$, obtained from the full electrostatic potential solution (full theory) and from the large gyro-orbit model for $\tau = 2$ and $\alpha = 3^{\circ}$ and $5^{\circ}$, are shown normalized to their peak value.
}
\label{fig-energyangle-tau2}
\end{figure}

The kinetic energy of an ion at the wall is $E = U - \Omega \phi_{\rm W} / B$ and the angle of impact of an ion with the wall surface is $\sin \theta = |v_x| / \sqrt{2E } $.
Thus, the components $v_z$ and $v_x$ of the ion velocity can be expressed as functions of $\bar{x}$, $E$ and $\theta$ via
\begin{align}
v_x  = - \sqrt{2E} \sin \theta    \rm ,
\end{align}
\begin{align} \label{vz-E-theta-xbar}
v_z = \sqrt{2\left( E - \chi_{\rm M} (\bar{x} ) + \frac{\Omega \phi_{\rm W} }{ B}  \right) } \rm .  
\end{align}
The energy-angle distribution $\zeta_{\rm W} (E, \theta)$ is calculated from $f_{\rm W} ( \vec{v} )$ using the equation
\begin{align} \label{energy-angle}
\zeta_{\rm W} (E,\theta ) = \int_{ \bar{x}_{\rm c}}^{\chi_{\rm M}^{-1} ( E + \Omega \phi_{\rm W} / B )}    \frac{\sqrt{2E}\cos \theta }{\sqrt{2\left( E  - \chi_{\rm M} (\bar{x} ) + \Omega \phi_{\rm W} / B \right) } } f_{\rm W} ( \vec{v} ) \Omega d \bar{x} \text{,}
%% f_{\rm W} ( \sqrt{2E} \sin \theta , \Omega \bar{x} , \sqrt{2\left( E  - \chi_{\rm M} (\bar{x} ) + \Omega \phi_{\rm W} / B \right) } ) 
\end{align}
where the Jacobian
\begin{align} \label{Jacobian}
 \frac{\partial (v_x, v_z )}{ \partial ( E, \theta ) } = \frac{\sqrt{2E}\cos \theta }{\sqrt{2\left( E  - \chi_{\rm M} (\bar{x} ) + \Omega \phi_{\rm W} / B \right) } } \rm 
\end{align}
was used to change variables from $v_x$ and $v_z$ to $E$ and $\theta$.
The inverse function of $\chi_{\rm M } (\bar{x})$, denoted $\chi_{\rm M}^{-1} $, is used to obtain the maximum value of $\bar{x}$ for a given value of $E$, which is, from equation (\ref{vz-E-theta-xbar}), the solution of $\chi_{\rm M} (\bar{x}) = E + \Omega \phi_{\rm W} / B $.

The energy-angle distributions calculated from the numerical solution of the electrostatic potential in the magnetic presheath and from the large gyro-orbit model are shown for $\alpha = 3^{\circ}$ and $5^{\circ}$, for $\tau = 0.5$ --- in figure~\ref{fig-energyangle-tau0pt5} --- and for $\tau =2$ --- in figure \ref{fig-energyangle-tau2}.
The qualitative features of the distribution function obtained from the full electrostatic potential solution are, even for $ \tau = 0.5$, adequately captured by the model, including the average angle of impact of ions with the wall.
The model performs better at the largest of the two values of $\tau$ ($\tau = 2 $, figure \ref{fig-energyangle-tau2}), as expected.

\subsection{Accounting for $\sqrt{Zm_{\rm e} / m_{\rm i}} \sim \alpha$} \label{subsec-VDMP-nonadiabatic}
 
For some of the angles we have considered, the assumption of adiabatic electrons, $\sqrt{Zm_{\rm e} / m_{\rm i} } = 0.0165 \approx 1^{\circ} \ll \alpha$, is not well-satisfied.
Once $ \phi_{\rm DSE} - \phi_{\rm W} \leqslant 0$ our assumption that the Debye sheath repels most electrons back into the magnetic presheath is clearly incorrect.
In fact, the Boltzmann distribution for the electron density becomes inaccurate when $\phi_{\rm DSE} - \phi_{\rm W}$ becomes sufficiently small that the ordering (\ref{phiDS}) is no longer satisfied.
The critical value of $\alpha$ for which $\phi_{\rm DSE} = \phi_{\rm W}$ in the model increases slightly with $\tau$: for $\tau = 2$ it is  $\alpha \approx 3^{\circ}$, while for $\tau = 10$ it is $\alpha \approx 5^{\circ}$. 
In order to solve for the self-consistent electrostatic potential across the magnetic presheath, a more accurate expression for the electron density must be used.
In the context of the large gyro-orbit model, this is expected to change equations (\ref{quasi-DSE}) and (\ref{Bohm}).

\section{Conclusions} \label{sec-conclusions}

The velocity distribution of ions reaching a planar target when the angle between the magnetic field and the target is small, $\alpha \ll 1$, was calculated using a model consisting of the set of equations (\ref{vy-DS}), (\ref{vz-DS}), (\ref{phi-wall}), (\ref{quasi-DSE}), (\ref{Bohm}), (\ref{mu-chiM-large}), (\ref{xbarc-model}) and (\ref{f-large-W}) (replaced with (\ref{f-large-DSE}) at the Debye sheath entrance instead of the target).
The model, like the asymptotic theory it is based on, was argued to be valid for $\alpha \leqslant 5^{\circ}$.
The advantage of the model is that the full solution of the quasineutrality equation in the magnetic presheath is bypassed, and replaced with constraints derived from quasineutrality near the Debye sheath entrance only.
The treatment is more accurate for large ion gyro-orbits, $\tau = T_{\rm i} / ZT_{\rm e} \gg 1 $.
Yet, it can be used also for $\tau \sim 1$ and reproduces the main qualitative features of distribution functions obtained by solving the self-consistent electrostatic potential across the magnetic presheath (for $\alpha \ll 1$), as shown in figures \ref{fig-thinf}, \ref{fig-energyangle-tau0pt5} and \ref{fig-energyangle-tau2}.
Since the sputtering yield of an ion striking a target depends on the ion's energy and angle of incidence with the target, calculations of energy-angle distributions (\ref{energy-angle}) using the model, shown in figures \ref{fig-energyangle-tau0pt5} and \ref{fig-energyangle-tau2}, may be valuable for sputtering predictions.

The narrowing of the wall-normal velocity distribution with the angle $\alpha$, shown in figure~\ref{fig-thinf} at the Debye sheath entrance, is explained from the model as follows.
Ions reaching the Debye sheath have a minimum normal velocity, $V_{x,\rm slow}$, which is related to the size of the gyro-orbit, and so to the adiabatic invariant $\mu$.
Ions with smaller gyro-orbits have a smaller gyration velocity, and so a smaller magnetic force acts on them to maintain the gyro-motion.
Consequently, a weaker electric force is needed to overcome the magnetic force and accelerate these ions towards the target.
Ions in smaller gyro-orbits (smaller $\mu$) are thus accelerated towards the wall for a larger distance, as shown schematically in figure \ref{fig-ion}(b-c).
However, the dependence of $V_{x,\rm slow}$ on the adiabatic invariant $\mu$ is weak, as seen in figure \ref{fig-Vxslow}.
Since the distribution function exponentially decays with $\mu$, the distribution function sharply drops to zero for $|v_x |$ below the typical values of $V_{x,\rm slow}$, as seen in figure~\ref{fig-thinf}.
The width of the wall-normal velocity distribution is therefore dominated by the gyrophase dependence of $v_x$ at the target.
This dependence is represented, schematically, by pairs of ion trajectories with the same value of $\mu$ and $U$ in figures \ref{fig-ion}(b-c).
It results in the scaling $\langle \tilde{v}_x^2 \rangle \sim \alpha v_{\rm t,i}^2$ for the variance of $v_x$.

The orderings (\ref{ordering-scales}) and (\ref{ordering-angle}) are required in the asymptotic theory and in the large gyro-orbit model, and are typically well-satisfied in fusion devices except for $\alpha \gg \sqrt{Zm_{\rm e} / m_{\rm i}}$ and $\rho_{\rm e} \ll \lambda_{\rm D} $.
Therefore, a kinetic model (instead of an adiabatic model) for the electrons should be used in the quasineutrality equation for $\phi(x)$ in the magnetic presheath.
This would change the electron contribution to the closure equations (\ref{quasi-DSE}) (quasineutrality) and (\ref{Bohm}) (kinetic Bohm condition) of the large gyro-orbit model.

~~

The author is grateful to Felix Parra for stimulating discussions and feedback. This work was supported by the US Department of Energy through grant DE-FG02-93ER-54197.

\appendix

\section{Change of $U_{\perp} - \chi_{\text{M}} (\bar{x})$ during the last ion gyro-orbit} \label{app-open}

In this appendix the change in the quantity $U_{\perp} - \chi_{\text{M}} (\bar{x})$ during the last gyro-orbit of an ion is calculated.
This quantity is denoted $\Delta_{\rm M} (\bar{x}, U )$, and is responsible for the spread of values of $v_x$ in the ion distribution function at the Debye sheath entrance (\ref{fopen-DSE}) and at the wall (\ref{fopen-W}).

Recalling from the discussion after equation (4.8) that $\dot{\bar{x}} = - \alpha v_z $, we obtain $\dot{\chi}_{\rm M} (\bar{x}) = \dot{x}_{\rm M}  \chi'(x_{\rm M}, \bar{x} ) + \dot{\bar{x}} \partial \chi (x_{\rm M}, \bar{x} ) / \partial \bar{x} = \alpha v_z \Omega^2 (x_{\rm M} - \bar{x} )$.
Here we have used that $\dot{x}_{\rm M}  \chi'(x_{\rm M}, \bar{x} ) = 0$ due to $\dot{x}_{\rm M} = 0$ for type I orbits ($x_{\rm M} = 0$) and $\chi' (x_{\rm M} , \bar{x}) = 0$ for type II orbits.
Also recalling $\dot{U}_{\perp} = - \alpha \Omega v_z v_y  = \alpha \Omega^2 (x - \bar{x})$, the rate of change of the quantity $U_{\perp} - \chi_{\text{M}} (\bar{x})$ is
\begin{align} \label{DeltaM-rateofchange}
\frac{d}{dt} \left( U_{\perp} - \chi_{\text{M}} (\bar{x}) \right) = \alpha \Omega^2 V_{\parallel} (\chi_{\rm M} (\bar{x} ), U ) \left( x - x_{\text{M}} \right) \text{.}
\end{align}
This is always positive for closed orbits which satisfy $x \geqslant x_{\rm b} \geqslant x_{\rm M}$.
Consider an ion, at a position $x$, that has just reached values of $\bar{x}$ and $U_{\perp}$ such that $U_{\perp} = \chi_{\rm M}(\bar{x}) $.
The time taken for the ion to reach $x=0$ is approximated by integrating the equation $dx/dt = v_x \simeq \sigma_x \sqrt{2\left( \chi_{\rm M} - \chi (x, \bar{x} ) \right)}$, where $\sigma_x = \pm 1$ is the sign of $v_x$, to get
\begin{align} \label{t-approx}
t = \left(\sigma_x +1 \right) \int_{x}^{x_{\rm t}} \frac{ds}{\sqrt{2\left( \chi_{\rm M} - \chi (s, \bar{x} ) \right)} } +  \int_0^{x} \frac{ds}{\sqrt{2\left( \chi_{\rm M} - \chi (s, \bar{x} ) \right)} }    \text{.}
\end{align}
Here, and in the rest of this section, we denote the position $x$ by the symbol $s$ when under an integral if the symbol $x$ is already used for one of the limits of the integration.
The problem with the approximation in (\ref{t-approx}) is that the second integral is logarithmically divergent for type II orbits due to the form of the integrand for $s \rightarrow x_{\rm M}$, 
\begin{align} \label{divergence}
\lim_{s\rightarrow x_{\rm M}} \frac{1}{\sqrt{2\left( \chi_{\rm M} - \chi (s, \bar{x} ) \right)}} = \frac{1}{\sqrt{\chi'' (x_{\rm M} )} \left| s- x_{\rm M} \right| } \text{.}
\end{align} 
However, the time $t$ taken by an ion to reach the Debye sheath entrance from a point in its last gyro-orbit would only be infinite if $v_x =  \sigma_x \sqrt{2\left( \chi_{\rm M} - \chi (x, \bar{x} ) \right)}$ was exactly true.
In practice, the quantity $U_{\perp} - \chi_{\rm M}(\bar{x})$ is not exactly zero.
To calculate this quantity, the time evolution of $U_{\perp} - \chi_{\rm M}(\bar{x})$ is estimated in the same way the time $t$ was estimated (incorrectly): we replace the time derivative in equation (\ref{DeltaM-rateofchange}) with a spatial derivative using the substitution $d/dt = v_x d/ dx$, and the approximation $v_x \simeq \sqrt{2\left( \chi_{\rm M} - \chi (x, \bar{x} ) \right)}$ to obtain
\begin{align}
\frac{d}{dx} \left( U_{\perp} - \chi_{\text{M}} \right) = \pm \alpha \Omega^2 V_{\parallel} \left( \chi_{\text{M}} (\bar{x}), U \right) \frac{ x - x_{\text{M}} }{ \sqrt{2\left( \chi_{\rm M} - \chi (x, \bar{x} ) \right)} } \text{.}
\end{align}
This equation is then integrated in the same way as before to obtain 
\begin{align} \label{int-Uperp-chiM-int}
U_{\perp} - \chi_{\text{M}}(\bar{x}) = & \alpha \Omega^2 V_{\parallel} \left( \chi_{\text{M}} (\bar{x}), U \right) \left[  \left(\sigma_x +1 \right) \int_{x}^{x_{\rm t}}  \frac{s - x_{\text{M}}}{\sqrt{2\left( \chi_{\text{M}} - \chi (s, \bar{x} ) \right) }} ds \right.  \nonumber \\
& \left. +  \int_0^{x}  \frac{s - x_{\text{M}}}{\sqrt{2\left( \chi_{\text{M}} - \chi (s, \bar{x} ) \right) }} ds  \right]  \text{.}
\end{align}
The second integral in (\ref{int-Uperp-chiM-int}) is not divergent near $x=x_{\rm M}$ because the integrand tends to 
\begin{align} \label{convergence}
\lim_{s\rightarrow x_{\rm M}} \frac{s-x_{\rm M}}{\sqrt{2\left( \chi_{\rm M} - \chi (s, \bar{x} ) \right)}} = \frac{s-x_{\rm M}}{\sqrt{\chi'' (x_{\rm M} )} \left| s- x_{\rm M} \right| } = \Theta \left( s-x_{\rm M} \right) \frac{1}{\sqrt{\chi'' (x_{\rm M} )}  } \text{,}
\end{align} 
which is always finite (moreover, the contribution from the region near $s=x_{\rm M}$ in the integral (\ref{int-Uperp-chiM-int}) vanishes because the integrand changes sign there).
Considering equation (\ref{int-Uperp-chiM-int}), $U_{\perp}$ is only ever exactly equal to $\chi_{\rm M} ( \bar{x}) $ at an instant, and at all other times it is different.
Therefore, the time estimated in (\ref{t-approx}) is incorrect, and the divergence in (\ref{divergence}) does not occur.
In practice, ions cross the effective potential maximum in a time $t \sim 2\pi |\ln \alpha | / \Omega$ \citep{Geraldini-2018}.

Upper and lower bounds for the values of $U_{\perp} - \chi_{\rm M}(\bar{x})$ of ions reaching $x=0$ can be obtained using the fact that these ions must have past trajectories with a bottom bounce point $x_{\rm b}$.
We consider the following two limiting cases: (i) an ion crossing the maximum $x=x_{\rm M}$ towards the sheath with $U_{\perp} = \chi_{\rm M}(\bar{x}) + \epsilon$; (ii) an ion bouncing back (for the last time) from $x=x_{\rm M}$ with $U_{\perp} = \chi_{\rm M}(\bar{x}) - \epsilon$, where $\epsilon$ is an energy difference so small it can be neglected.
The minimum value of $U_{\perp} - \chi_{\rm M}$ of an ion entering the Debye sheath is calculated from case (i),
\begin{align}
U_{\perp} - \chi_{\text{M}}(\bar{x}) = - \Delta_+ (x, \bar{x}, U )   \text{,}
\end{align}
where
\begin{align}
\Delta_+ (x, \bar{x}, U )  =  \alpha \Omega^2 V_{\parallel} \left( \chi_{\text{M}}, U \right)   \int_0^{x_{\rm M}}  \frac{ x_{\text{M}} - s}{\sqrt{2\left( \chi_{\text{M}} - \chi (s, \bar{x} ) \right) }} ds
\end{align} 
is a positive quantity.
Here, we have added to $U_{\perp} - \chi_{\text{M}}(\bar{x}) = 0$ the amount obtained by integrating equation (\ref{DeltaM-rateofchange}) from $x_{\rm M}$ to the Debye sheath entrance ($x\simeq 0$ here).
The maximum value of $U_{\perp} - \chi_{\rm M} (\bar{x})$ is calculated from case (ii),
\begin{align}
U_{\perp} - \chi_{\text{M}}(\bar{x}) = \Delta_{\rm M} (\bar{x}, U ) - \Delta_+ (x, \bar{x}, U )  \text{,}
\end{align}
where
\begin{align} \label{DeltaM-1}
\Delta_{\rm M} (\bar{x}, U ) = 2 \alpha \Omega^2 V_{\parallel} \left( \chi_{\text{M}}, U \right) \int_{x_{\rm M}}^{x_{\rm t}}  \frac{x - x_{\text{M}}}{\sqrt{2\left( \chi_{\text{M}} - \chi (x, \bar{x} ) \right) }} dx  \text{.}
\end{align}
Here, we have added to $U_{\perp} - \chi_{\text{M}}(\bar{x}) = 0$ the amount obtained by integrating equation (\ref{DeltaM-rateofchange}) from $x_{\rm M}$ to $x_{\rm t}$, then back again all the way to the Debye sheath entrance ($x\simeq 0$).
The quantity $\Delta_{+}$ was shown to be negligible when calculating $v_x$ from equation (\ref{vx-xbar-Uperp-x}), as it is always small relative to either $\chi_{\rm M} (\bar{x}) - \chi ( x, \bar{x})$ or $\Delta_{\rm M}(\bar{x}, U )$ \citep{Geraldini-2018}. 
Thus, we can consider $0 \leqslant U_{\perp} - \chi_{\rm M}(\bar{x}) < \Delta_{\rm M} (\bar{x}, U )$ for ions reaching the Debye sheath entrance.

Equation (\ref{DeltaM-def}) for $\Delta_{\rm M} (\bar{x}, U )$ follows from (\ref{DeltaM-1}) and from the equality
\begin{align}
\mu_{\rm op}'(\bar{x}) = \Omega^2 \int_{x_{\rm M}}^{x_{\rm t}}  \frac{x - x_{\text{M}}}{\sqrt{2\left( \chi_{\text{M}} - \chi (x, \bar{x} ) \right) }} dx \rm ,
\end{align}
which can be verified from (\ref{mu-op}).

\section{Ion conservation} \label{app-ionconservation}

The ion distribution function at the Debye sheath entrance, equation (\ref{fopen-DSE}), is proved here to be consistent with ion conservation in the magnetic presheath.
Equation (\ref{current-ion-parallel}) gives the current flowing normal to the wall at the magnetic presheath entrance.
In steady state, the current flowing normal to the wall at the Debye sheath entrance should be the same.
At the Debye sheath entrance, the ion density is small in $\alpha$ and the ion current flowing normal to the wall is due to the component $v_x$ of the the velocity of all ions,
\begin{align}
\frac{j_{\text{i},x}}{Ze} = & - 2\pi\int_{ \bar{x}_{\rm c}}^{\infty} \Omega d \bar{x} \int_{\Omega \mu}^{\infty}  \frac{ F \left( \mu_{\rm op} (\bar{x}) , U \right) dU}{ V_{\parallel } ( \chi_{\rm M} (\bar{x} ), U )  }   \\ & \times \int_{-\infty}^{\infty}  \hat{\Pi} \left(  \frac{1}{2}v_x^2 - \chi_{\rm M} ( \bar{x} ) + \frac{1}{2} \Omega^2 \bar{x}^2 + \frac{\Omega \phi_{\rm DSE}}{B}  ,  0 , \Delta_{\rm M}(\bar{x}, U )  \right) v_x dv_x
\end{align}
The last integral in $v_x$ is taken by replacing $v_x dv_x = d\left(v_x^2/2\right)$, and the result is $\Delta_{\rm M} (\bar{x}) = 2\alpha  \pi  \mu_{\rm op}'(\bar{x}) V_{\parallel } ( \chi_{\rm M} (\bar{x} ), U ) $,
\begin{align} \label{flow-0}
\frac{j_{\text{i},x}}{Ze} = & - 2 \alpha  \pi  \int_{0}^{\infty} \Omega d\bar{x} \mu_{\rm op}'(\bar{x}) \int_{\Omega \hat{\mu}}^{\infty} F \left(\mu_{\rm op}(\bar{x}) , U \right)  dU  \text{.} 
\end{align}
Using $ \mu_{\rm op}'(\bar{x}) = d\mu / d\bar{x}$ and changing integration variable to $\mu = \mu_{\rm op}(\bar{x})$ leads to equation (\ref{current-ion-parallel}).
The same argument applies to the ion distribution function at the wall, (\ref{fopen-W}), and to the large gyro-orbit model distribution functions, (\ref{f-large-DSE}) and (\ref{f-large-W}).

\bibliographystyle{jpp}

\bibliography{gyrokineticsbibliography}

\end{document}